\PassOptionsToPackage{table}{xcolor}
\documentclass[sigconf,nonacm]{acmart}
\usepackage{tabularx}
\usepackage{colortbl}
\usepackage{listings}
\usepackage{placeins}
\usepackage{stfloats}
\usepackage{tikz}
\newcommand{\CodeURL}{https://github.com/BITMinhaoSun/ISAOPD}
\newcommand{\CheckpointURL}{https://huggingface.co/sunminhao/ISAOPD-Qwen2.5-3B}
\newcommand{\WebsiteURL}{https://bitminhaosun.github.io/ISA-OPD/}

\definecolor{projectInk}{HTML}{233E53}
\definecolor{projectLine}{HTML}{CCD8E0}
\definecolor{projectTint}{HTML}{F3F7F9}

\tikzset{
  pics/source-code/.style={code={
    \draw (-2,2) -- (-4,0) -- (-2,-2) (2,2) -- (4,0) -- (2,-2)
          (.8,3) -- (-.8,-3);
  }},
  pics/checkpoint/.style={code={
    \node[inner sep=0pt,outer sep=0pt] at (0,0)
      {\includegraphics[width=11pt]{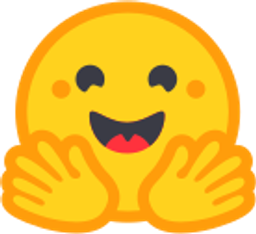}};
  }},
  pics/website/.style={code={
    \draw (0,0) circle (3.7) (0,0) ellipse (1.6 and 3.7) (-3.7,0) -- (3.7,0);
  }}
}

\newcommand{\projectBadge}[2]{%
  \begin{tikzpicture}[baseline=(label.base)]
    \node[rounded corners=7pt, draw=projectLine, fill=projectTint,
          text=projectInk, line width=.45pt, inner xsep=7pt, inner ysep=2.5pt,
          font=\normalfont\sffamily\small] (label) {\hspace*{11pt}\strut #2};
    \pic[x=1pt,y=1pt,draw=projectInk,line width=.6pt,line cap=round,line join=round]
      at ([xshift=8.5pt]label.west) {#1};
  \end{tikzpicture}%
}
\newcommand{\projectLink}[3]{%
  \ifdefempty{#1}{\projectBadge{#2}{#3}}{\href{#1}{\projectBadge{#2}{#3}}}%
}

\newsavebox{\projectResourcesBox}

\makeatletter
\AddToHook{cmd/@mkauthors/after}{%
  \setbox\projectResourcesBox=\vbox{%
    \hbox to \textwidth{\hfil
      \projectLink{\WebsiteURL}{website}{Project Page}\hspace{5pt}%
      \projectLink{\CheckpointURL}{checkpoint}{Model Checkpoints}\hspace{5pt}%
      \projectLink{\CodeURL}{source-code}{Code}%
      \hfil}%
    \vskip 6pt
  }%
  \setbox\mktitle@bx=\vbox{%
    \vskip-\ht\projectResourcesBox
    \nointerlineskip\unvbox\mktitle@bx
    \nointerlineskip\box\projectResourcesBox
  }%
}
\makeatother

\setcopyright{none}
\acmDOI{}
\acmISBN{}
\begin{document}
\raggedbottom
\title[Same Trajectory, Different Knowledge]{Same Trajectory, Different Knowledge:\texorpdfstring{\\}{ }Information-State-Aware On-Policy Distillation for Search Agents}

\author{Minhao Sun}
\affiliation{%
  \institution{Beijing Institute of Technology}
  \city{Beijing}
  \country{China}}
\author{Xian-Ling Mao}
\authornote{Corresponding author.}
\affiliation{%
  \institution{Beijing Institute of Technology}
  \city{Beijing}
  \country{China}}
\author{Eric Tao}
\affiliation{%
  \institution{Huawei}
  \city{Beijing}
  \country{China}}
\author{Zi-Ao Ma}
\affiliation{%
  \institution{Beijing Institute of Technology}
  \city{Beijing}
  \country{China}}
\author{Zhijing Wu}
\affiliation{%
  \institution{Beijing Institute of Technology}
  \city{Beijing}
  \country{China}}
\renewcommand{\shortauthors}{Minhao Sun et al.}

\begin{abstract}
On-policy distillation trains search agents with teacher feedback on student-generated trajectories. Recent methods augment the teacher’s context with supporting evidence, reference answers, or hindsight information. However, this privileged information may allow the teacher to rely on facts that the student cannot reliably recover from its current state. The resulting information-state mismatch can lead the teacher to undervalue retrieval that helps the student establish those facts. To address this
mismatch, we introduce Information-State-Aware Distillation (ISA-Distill), which adapts teacher feedback to the student’s factual recoverability. ISA-Distill estimates how reliably the student can recover action-relevant knowledge and derives teacher feedback for the same sampled action from two complementary views. The fact-use view supplies the teacher with the relevant fact, producing feedback that guides the student in using knowledge. The evidence-seeking view instead presents the information need, producing feedback that guides the student in acquiring supporting evidence. The two signals are weighted toward fact use when recoverability is high and toward evidence seeking when it is low, and the resulting feedback guides policy optimization. Experiments across diverse QA benchmarks and model scales demonstrate improved answer accuracy and retrieval efficiency.
\end{abstract}
\maketitle

\section{Introduction}
Search agents support Web information access by interleaving language-model reasoning with external retrieval to answer knowledge-intensive questions~\cite{yao2023react,tao2026race}. Through retrieval, they can establish intermediate facts needed for subsequent reasoning. The need for retrieval depends on how reliably the agent can already recover these facts from its current context and internal knowledge. Learning an effective search policy therefore calls for feedback that reflects the agent's information needs at each step.

On-policy distillation (OPD) uses teacher feedback to train such a policy on student-generated trajectories~\cite{ross2011dagger,agarwal2024gkd}, placing supervision at states the learner actually visits. However, teacher feedback on student-generated trajectories does not guarantee that the student has access to the same knowledge as the teacher. During supervision, the teacher may receive privileged information unavailable to the student, such as supporting passages, reference answers, or information from other rollouts~\cite{xie2026evisd,ma2026sdsearch}.

\begin{figure}[t]
    \centering
    \includegraphics[width=\columnwidth]{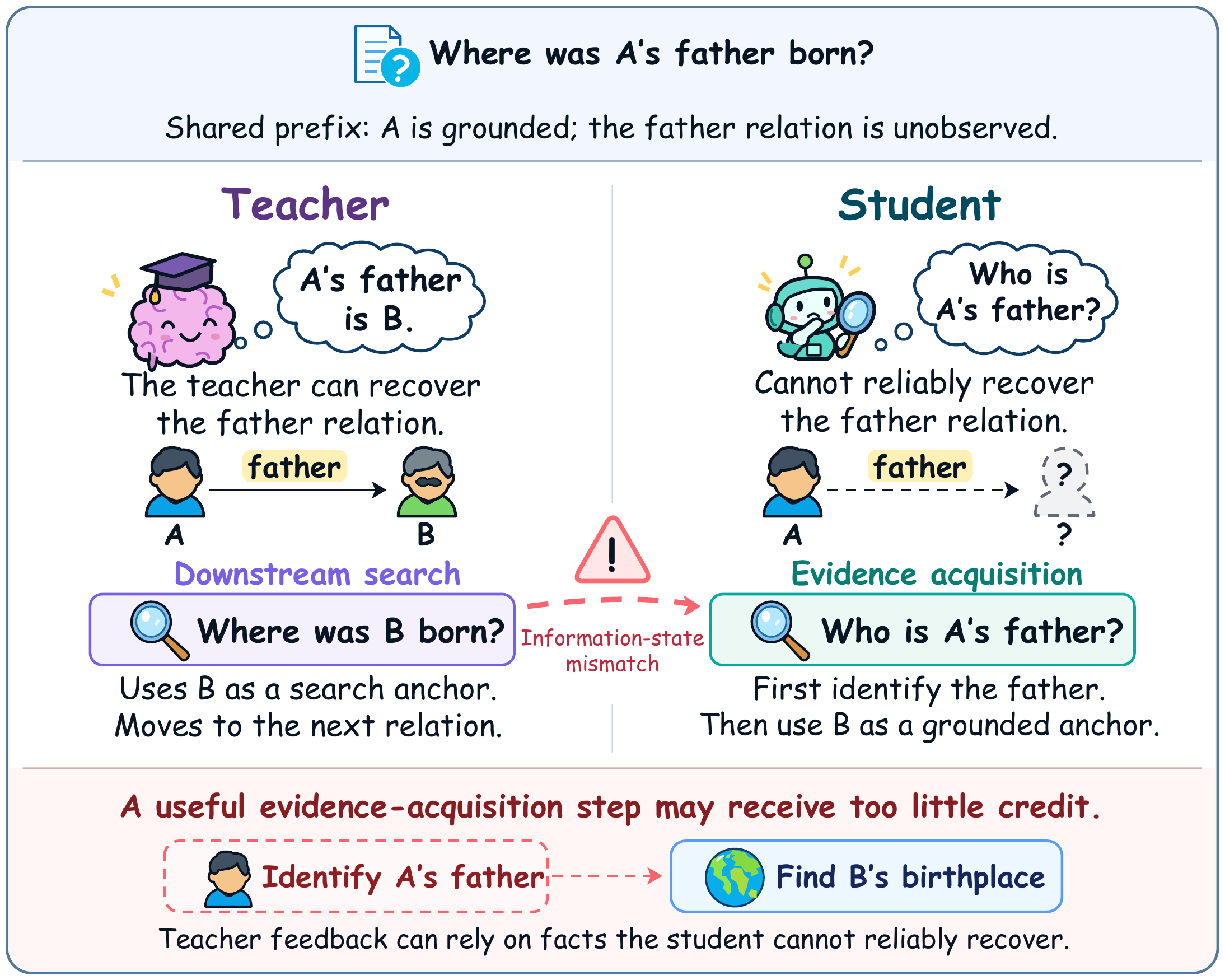}
    \caption{Information-state mismatch at a shared trajectory prefix. A teacher that can recover an intermediate fact may favor using it directly and undervalue a search that helps the student establish it.}
    \Description{For the question Where was A's father born, the teacher knows that A's father is B through parametric knowledge or privileged information. The student has not observed this relation and cannot reliably recover B. The illustration contrasts a query about B's birthplace with a query that identifies A's father. The latter is one possible way to establish the missing relation and may receive too little credit from feedback that assumes the relation is already available.}
    \label{fig:information-state-mismatch}
\end{figure}

This \emph{information-state mismatch} can lead the teacher to favor actions that rely on facts available in its privileged context and undervalue searches that the student needs to establish those facts, as illustrated in Figure~\ref{fig:information-state-mismatch}. Teacher feedback should therefore reflect whether the student needs help establishing a fact or using it. When the student cannot reliably recover a needed fact, feedback should give credit to searches that help establish it. When the student can recover the fact, feedback can place greater emphasis on using it effectively. Balancing these two forms of guidance requires accounting for the student's current information state. Specifically, we assess how likely the student is to correctly recover an action-relevant fact from its current context and internal knowledge without further retrieval. This \emph{factual recoverability} can change as the student acquires new evidence.

Building on this principle, we introduce Information-State-Aware Distillation (ISA-Distill), which adapts teacher feedback to the student’s factual recoverability. For a focal fact selected for a sampled action, ISA-Distill estimates recoverability from the student’s responses to a value-masked probe of its pre-step state. It re-scores the same action under two complementary teacher views: a fact-use view that makes the focal fact available, and an evidence-seeking view that masks its value in the added context and emphasizes evidence acquisition. The recoverability estimate determines the relative weight of the two views, and their combined feedback provides a bounded correction to outcome-based credit~\cite{xie2026evisd}.
In this way, ISA-Distill guides the student to use facts it can reliably recover and retrieve evidence for those it cannot.
We summarize our main contributions as follows:
\begin{itemize}
    \item \textbf{Identifying information-state mismatch.} We identify information-state mismatch in on-policy distillation and provide paired evidence that supplying a fact can make teacher feedback less favorable toward searches with positive observed evidence utility.
    \item \textbf{Recoverability-guided distillation.} We propose ISA-Distill, which incorporates pre-step factual recoverability into on-policy distillation to balance fact-use and evidence-seeking feedback according to the student's current state.
    \item \textbf{Improved accuracy and search efficiency.} Across seven QA benchmarks, ISA-Distill achieves higher macro-average EM than the evaluated baselines through self-distillation at both 3B and 7B (Tables~\ref{tab:main-7b} and~\ref{tab:main-3b}). In a controlled comparison with EviSD, it makes more searches on multi-hop tasks and fewer on single-hop tasks, with fewer calls overall, as shown in Figure~\ref{fig:search-allocation-7b}.
\end{itemize}

\section{Related Work}
\subsection{On-Policy Distillation with Privileged Information}
DAgger collects expert feedback at learner-visited states~\cite{ross2011dagger}, and GKD trains language models with teacher feedback on self-generated outputs~\cite{agarwal2024gkd}. Privileged-information methods augment teacher inputs with context unavailable to students at inference~\cite{penaloza2026privileged}. LEAP examines the trade-off between better-informed expert guidance and what students can learn to reproduce from their own observations~\cite{choudhury2024leap}. DOPD routes token-level supervision between privileged teacher and student policies using their advantage gap and relative probabilities~\cite{yu2026dopd}. AR-OPD combines a partially privileged anchor with residual guidance from a fully privileged view to limit locally unsupported imitation~\cite{zhang2026aropd}. PSP supplies privileged instructions during rollout sampling to guide student practice~\cite{su2026psp}. ISA-Distill focuses on aligning privileged teacher feedback with the student's current information state.

\subsection{Credit Assignment for Search Agents}
Search-R1 learns reasoning and retrieval through reinforcement learning (RL) with final-answer rewards~\cite{jin2025searchr1}. IGPO and TIPS provide local credit through changes in student and teacher correct-answer likelihoods, respectively~\cite{wang2026igpo,xie2026tips}; GiGPO estimates step-level relative advantages by grouping actions from repeated states~\cite{feng2025gigpo}. HiPRAG uses knowledge-grounded process rewards to assess search necessity~\cite{wu2026hiprag}. In knowledge-base question answering, GAPD aligns student and gold execution states through explored entities and conditions token-level guidance on the aligned gold action~\cite{sun2026gapd}. SD-Search uses rollout-group queries and correctness labels as hindsight context for feedback on search-query tokens~\cite{ma2026sdsearch}. EviSD re-scores search actions with supporting evidence and answer actions with gold answers, converting token-level contrasts into bounded adjustments to outcome-based advantages~\cite{xie2026evisd}. ISA-Distill refines outcome-based credit by combining fact-use and evidence-seeking feedback on the same sampled action.

\subsection{Knowledge Estimation for Adaptive Feedback}
Aichberger et al. estimate uncertainty from the likelihood of a single greedily decoded response~\cite{aichberger2026rethinking}. Chen et al. distinguish mastered, confused, and missing knowledge to guide targeted augmentation and improve confidence calibration~\cite{chen2026know}. KbSD uses question-level internal reliability and retrieval sufficiency to construct teacher hints and select quadrant-specific objectives~\cite{feng2026kbsd}. Within a search trajectory, access to individual facts can change as evidence accumulates. ISA-Distill estimates the recoverability of an action-relevant fact from the student's pre-step state without further retrieval, providing a local signal for adapting teacher feedback.

\begin{figure}[t]
    \centering
    \includegraphics[width=\columnwidth]{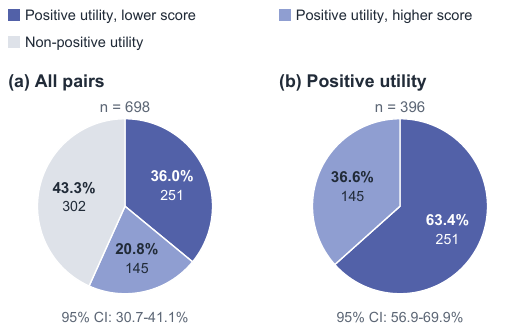}
    \caption{Teacher-score changes after adding the focal fact: (a) all pairs; (b) positive-utility pairs. Labels show percentages and counts; 95\% cluster-bootstrap CIs refer to the dark segments.}
    \Description{Two pie charts share the same categories. Positive utility means that student final-answer EM is higher with intact focal support than with that support masked; non-positive utility means equal or lower EM. Lower and higher scores compare the teacher's mean token log-probability for the same search with and without the added focal fact. Across all 698 pairs, 251 have positive observed evidence utility and lower teacher scores, 145 have positive utility and higher scores, and 302 have non-positive utility regardless of score change. Among the 396 positive-utility pairs, 63.4 percent receive lower scores and 36.6 percent receive higher scores.}
    \label{fig:paired-diagnostic}
\end{figure}

\begin{figure*}[t]
    \centering
    \includegraphics[width=\textwidth]{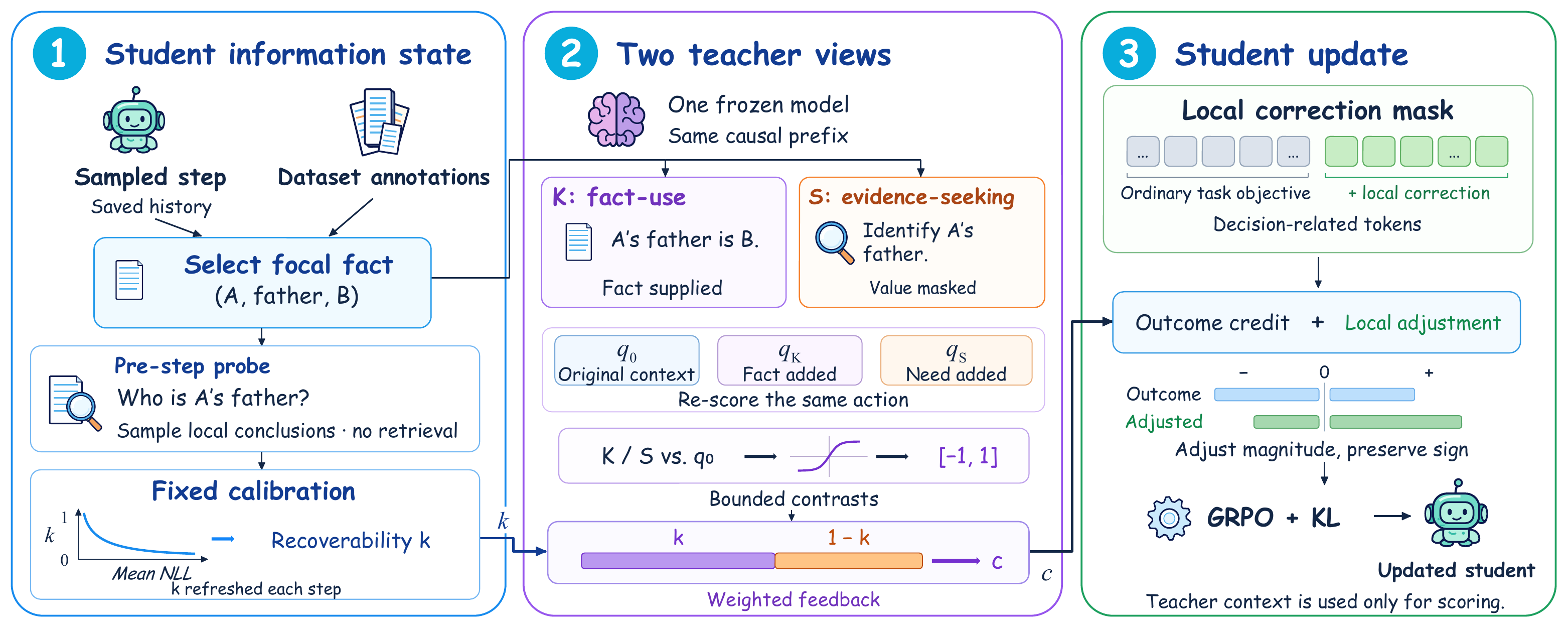}
    \caption{The recoverability estimate $k$ weights fact-use ($K$) and evidence-seeking ($S$) feedback on the same student action. Both views use the frozen rollout policy in self-distillation; their weighted feedback adjusts outcome-based credit.}
    \label{fig:framework}
    \Description{The selector uses dataset annotations to select an action-relevant focal fact and formulate a value-masked probe. The frozen student samples local conclusions from the saved pre-step state and probe. Their mean NLL $u$ is mapped to a recoverability estimate $k$, which weights fact-use and evidence-seeking feedback on the original action. The fact-use view receives the focal fact; the evidence-seeking view receives the same information need. Their combined feedback adjusts outcome-based credit.}
\end{figure*}

\definecolor{baselinegroupblue}{RGB}{225,237,250}
\setcounter{dbltopnumber}{2}

\begin{table*}[t]
\centering
\caption{Main results on seven QA benchmarks with Qwen2.5-7B-Instruct. Dataset columns report EM (\%). Avg. is the macro-average EM, with equal weight for each of the seven datasets. Bold and underlining mark the highest and second-highest displayed values in each column, including ties.}
\label{tab:main-7b}
\begingroup
\fontsize{8.5}{9.6}\selectfont
\setlength{\tabcolsep}{2.8pt}
\renewcommand{\arraystretch}{1.0}
\begin{tabularx}{\textwidth}{@{}p{70pt}*{8}{>{\raggedleft\arraybackslash}X}@{}}
\toprule
& \multicolumn{3}{c}{Single-hop QA} & \multicolumn{4}{c}{Multi-hop QA} & \multicolumn{1}{c}{Overall} \\
\cmidrule(lr){2-4}\cmidrule(lr){5-8}\cmidrule(lr){9-9}
\textbf{Method} & \textbf{NQ} & \textbf{TriviaQA} & \textbf{PopQA} & \textbf{HotpotQA} & \textbf{2WikiMQA} & \textbf{MuSiQue} & \textbf{Bamboogle} & \textbf{Avg.} $\uparrow$ \\
\midrule
\rowcolor{baselinegroupblue}
\multicolumn{9}{@{\hspace{3pt}}l@{}}{\fontsize{8}{9.2}\selectfont\color{black!80}\normalfont\bfseries Outcome-Based RL and Reward Shaping} \\
Search-R1~\cite{jin2025searchr1} & 42.9 & 62.3 & 42.7 & 38.6 & 34.6 & 16.2 & 40.0 & 39.6 \\
AutoRefine~\cite{shi2025autorefine} & 42.0 & 60.4 & 41.3 & 39.8 & 33.3 & 17.6 & 39.2 & 39.1 \\
\addlinespace[1.5pt]
\rowcolor{baselinegroupblue}
\multicolumn{9}{@{\hspace{3pt}}l@{}}{\fontsize{8}{9.2}\selectfont\color{black!80}\normalfont\bfseries Turn-/Step-Level Credit Assignment} \\
TIPS~\cite{xie2026tips} & 43.4 & 64.3 & 44.5 & 43.0 & 43.0 & 17.1 & 36.8 & 41.7 \\
PiCA~\cite{liu2026pica} & 46.0 & 64.1 & 44.2 & 42.4 & 40.1 & 19.7 & 41.9 & 42.6 \\
GiGPO~\cite{feng2025gigpo} & 46.4 & 64.7 & 46.1 & 41.6 & 43.6 & 18.9 & \textbf{68.9} & 47.2 \\
\addlinespace[1.5pt]
\rowcolor{baselinegroupblue}
\multicolumn{9}{@{\hspace{3pt}}l@{}}{\fontsize{8}{9.2}\selectfont\color{black!80}\normalfont\bfseries Self-Distillation: Rollout-Conditioned Privileged Information} \\
SD-Search~\cite{ma2026sdsearch} & \underline{49.5} & \underline{66.8} & \textbf{49.4} & \underline{47.1} & 44.1 & \underline{22.2} & \underline{54.4} & \underline{47.6} \\
\addlinespace[1.5pt]
\rowcolor{baselinegroupblue}
\multicolumn{9}{@{\hspace{3pt}}l@{}}{\fontsize{8}{9.2}\selectfont\color{black!80}\normalfont\bfseries Self-Distillation: Skill-Conditioned Privileged Information} \\
Skill-SD~\cite{wang2026skillsd} & 47.1 & 64.5 & 44.2 & \textbf{47.8} & 42.1 & 20.2 & 38.4 & 43.5 \\
RLSD$_{\mathrm{Skill}}$~\cite{yang2026rlsd} & 46.8 & 63.0 & 44.4 & 45.5 & \underline{48.9} & 21.5 & 46.4 & 45.2 \\
SDAR~\cite{lu2026sdar} & 46.3 & 63.5 & 48.2 & 43.8 & 48.4 & 19.6 & 46.4 & 45.2 \\
\addlinespace[1.5pt]
\rowcolor{baselinegroupblue}
\multicolumn{9}{@{\hspace{3pt}}l@{}}{\fontsize{8}{9.2}\selectfont\color{black!80}\normalfont\bfseries Self-Distillation: Answer-Conditioned Privileged Information} \\
RLSD$_{\mathrm{Ans}}$~\cite{yang2026rlsd} & 46.7 & 64.2 & 45.5 & 41.8 & 42.9 & 16.6 & 32.0 & 41.4 \\
\addlinespace[1.5pt]
\rowcolor{baselinegroupblue}
\multicolumn{9}{@{\hspace{3pt}}l@{}}{\fontsize{8}{9.2}\selectfont\color{black!80}\normalfont\bfseries Self-Distillation: Instance-Level Evidence} \\
EviSD~\cite{xie2026evisd} & 48.9 & 66.5 & 48.6 & 46.6 & 48.6 & 21.6 & 49.6 & 47.2 \\
\addlinespace[2pt]
\midrule
\textbf{ISA-Distill (ours)} & \textbf{50.3} & \textbf{67.1} & \underline{48.8} & \underline{47.1} & \textbf{49.9} & \textbf{22.4} & 50.4 & \textbf{48.0} \\
\bottomrule
\end{tabularx}
\endgroup
\end{table*}

\newcommand{\MainThreeBResults}{%
\begin{table*}[t]
\centering
\caption{Main results on seven QA benchmarks with Qwen2.5-3B-Instruct. Dataset columns report EM (\%). Avg. is the macro-average EM, with equal weight for each of the seven datasets. Bold and underlining mark the highest and second-highest displayed values in each column, including ties.}
\label{tab:main-3b}
\begingroup
\fontsize{8.5}{9.6}\selectfont
\setlength{\tabcolsep}{2.8pt}
\renewcommand{\arraystretch}{1.0}
\begin{tabularx}{\textwidth}{@{}p{70pt}*{8}{>{\raggedleft\arraybackslash}X}@{}}
\toprule
& \multicolumn{3}{c}{Single-hop QA} & \multicolumn{4}{c}{Multi-hop QA} & \multicolumn{1}{c}{Overall} \\
\cmidrule(lr){2-4}\cmidrule(lr){5-8}\cmidrule(lr){9-9}
\textbf{Method} & \textbf{NQ} & \textbf{TriviaQA} & \textbf{PopQA} & \textbf{HotpotQA} & \textbf{2WikiMQA} & \textbf{MuSiQue} & \textbf{Bamboogle} & \textbf{Avg.} $\uparrow$ \\
\midrule
\rowcolor{baselinegroupblue}
\multicolumn{9}{@{\hspace{3pt}}l@{}}{\fontsize{8}{9.2}\selectfont\color{black!80}\normalfont\bfseries Outcome-Based RL and Reward Shaping} \\
Search-R1~\cite{jin2025searchr1} & 39.7 & 56.5 & 39.1 & 33.1 & 31.0 & 12.4 & 23.2 & 33.6 \\
AutoRefine~\cite{shi2025autorefine} & 43.6 & 59.7 & 44.7 & 40.4 & 38.0 & 16.9 & 33.6 & 39.6 \\
\addlinespace[1.5pt]
\rowcolor{baselinegroupblue}
\multicolumn{9}{@{\hspace{3pt}}l@{}}{\fontsize{8}{9.2}\selectfont\color{black!80}\normalfont\bfseries Turn-/Step-Level Credit Assignment} \\
TIPS~\cite{xie2026tips} & 43.5 & 58.8 & 42.8 & 31.4 & 29.3 & 8.7 & 20.8 & 33.6 \\
PiCA~\cite{liu2026pica} & 42.6 & 61.2 & 41.7 & 40.0 & 40.8 & 16.0 & 34.7 & 39.6 \\
GiGPO~\cite{feng2025gigpo} & 42.0 & 59.5 & 42.4 & 36.9 & 37.0 & 12.6 & 28.8 & 37.0 \\
\addlinespace[1.5pt]
\rowcolor{baselinegroupblue}
\multicolumn{9}{@{\hspace{3pt}}l@{}}{\fontsize{8}{9.2}\selectfont\color{black!80}\normalfont\bfseries Self-Distillation: Rollout-Conditioned Privileged Information} \\
SD-Search~\cite{ma2026sdsearch} & \underline{46.9} & 62.4 & 46.5 & \textbf{42.4} & 41.9 & \underline{18.4} & \textbf{40.4} & \underline{42.7} \\
\addlinespace[1.5pt]
\rowcolor{baselinegroupblue}
\multicolumn{9}{@{\hspace{3pt}}l@{}}{\fontsize{8}{9.2}\selectfont\color{black!80}\normalfont\bfseries Self-Distillation: Skill-Conditioned Privileged Information} \\
Skill-SD~\cite{wang2026skillsd} & 44.4 & 60.4 & 44.0 & 39.5 & 40.4 & 15.4 & 30.4 & 39.2 \\
RLSD$_{\mathrm{Skill}}$~\cite{yang2026rlsd} & 41.5 & 58.6 & 42.3 & 40.4 & 40.2 & 16.8 & 34.4 & 39.2 \\
SDAR~\cite{lu2026sdar} & 44.8 & 58.1 & 44.3 & 38.6 & 36.2 & 15.7 & 32.8 & 38.6 \\
\addlinespace[1.5pt]
\rowcolor{baselinegroupblue}
\multicolumn{9}{@{\hspace{3pt}}l@{}}{\fontsize{8}{9.2}\selectfont\color{black!80}\normalfont\bfseries Self-Distillation: Answer-Conditioned Privileged Information} \\
RLSD$_{\mathrm{Ans}}$~\cite{yang2026rlsd} & 44.8 & 61.8 & 47.5 & 40.1 & 41.3 & 14.6 & 28.8 & 39.8 \\
\addlinespace[1.5pt]
\rowcolor{baselinegroupblue}
\multicolumn{9}{@{\hspace{3pt}}l@{}}{\fontsize{8}{9.2}\selectfont\color{black!80}\normalfont\bfseries Self-Distillation: Instance-Level Evidence} \\
EviSD~\cite{xie2026evisd} & 46.6 & \underline{63.1} & \textbf{48.7} & \textbf{42.4} & \underline{44.9} & 17.5 & 35.2 & 42.6 \\
\addlinespace[2pt]
\midrule
\textbf{ISA-Distill (ours)} & \textbf{47.6} & \textbf{63.3} & \underline{47.7} & \underline{42.0} & \textbf{45.9} & \textbf{20.5} & \underline{40.0} & \textbf{43.9} \\
\bottomrule
\end{tabularx}
\endgroup
\end{table*}
}

\section{Background and Motivation}
\label{sec:preliminary}

\subsection{On-Policy Distillation}
\label{sec:opd-preliminary}
OPD trains a student policy on student-generated trajectories using teacher feedback, with reverse KL as a common objective~\cite{agarwal2024gkd}. At a student-visible prefix $\xi_n$, the frozen rollout policy samples $x_n\sim\pi_{\mathrm{old}}(\cdot\mid\xi_n)$. The teacher distribution is $q_{T,n}=\pi_T(\cdot\mid D_n\oplus\xi_n)$, where $D_n$ is training-only privileged context. With the rollout policy and teacher fixed within a batch, the sampled-token contrast is
\begin{equation}
\delta_n=\log\frac{q_{T,n}(x_n)}{\pi_{\mathrm{old}}(x_n\mid\xi_n)}.
\label{eq:opd-token-contrast}
\end{equation}
Positive $\delta_n$ means that the teacher assigns the sampled token a higher probability than the rollout policy.

Even when the teacher and rollout policy share the same frozen model, privileged context can change teacher scores for the same sampled search action.

\subsection{Lower Teacher Scores for Useful Search}
\label{sec:paired-diagnostic}
We examine 698 archived state--search-action pairs from HotpotQA~\cite{yang2018hotpotqa} training questions, using a frozen 3B checkpoint for teacher scoring and student continuations. The teacher score is the mean log-probability of the original search tokens, evaluated with and without an added reference-supported focal fact while holding all other context and instructions fixed. Student evidence utility is the difference in final-answer EM between intact retrieval results and results with only focal-support spans masked. Pairs with non-positive utility are grouped regardless of score change.

As shown in Figure~\ref{fig:paired-diagnostic}, supplying the focal fact lowers teacher scores for 251 of 396 searches with positive observed evidence utility (63.4\%; 95\% CI: 56.9--69.9\%). The sampling protocol and additional controls are specified in Appendix~\ref{app:paired-diagnostic}.

\section{Method}
\label{sec:method}
As illustrated in Figure~\ref{fig:framework}, ISA-Distill selects an action-relevant fact, estimates its recoverability from the student's pre-step state, and uses this estimate to weight two teacher views of the same sampled action. The combined feedback adjusts outcome-based credit during policy optimization.

\subsection{State and Focal-Fact Construction}
\label{sec:problem-setting}
Before step $t$, the student state is
\begin{equation}
s_t^-=(q,h_{<t},E_{<t},b_t),
\label{eq:preaction-state}
\end{equation}
where $q$ is the question, $h_{<t}$ is prior student history, $E_{<t}$ contains completed observations, and $b_t$ is the remaining search budget. Given this state and the sampled action, a selector chooses an action-relevant focal fact from training-only supporting evidence and gold answers. It constructs a value-masked probe using entities and relations in the visible context, omitting the target value and its reference-only aliases (Appendix~\ref{app:focus-details}).

\subsection{Estimating Factual Recoverability}
\label{sec:boundary-estimation}
\label{sec:likelihood-scoring}
\label{sec:probe-validation}
To estimate recoverability before the action, we form $P$ from $s_t^-$ and the probe, excluding the current step's reasoning, query, and retrieval results. The frozen rollout policy $\pi=\pi_{\mathrm{old}}$ samples 32 local conclusions $\tilde y^{(1)},\ldots,\tilde y^{(32)}$ from $P$ without further retrieval. For a conclusion $y=(y_1,\ldots,y_M)$, excluding padding, mean token negative log-likelihood (NLL) is
\[
U_{\pi}(y\mid P)=-\frac{1}{M}\sum_{m=1}^{M}
\log\pi(y_m\mid P,y_{<m}).
\]
Lower NLL indicates stronger token-level support~\cite{aichberger2026rethinking} and is associated with question-level correctness~\cite{chen2026know}. Across 50,000 Qwen2.5-7B-Instruct~\cite{qwen2024qwen25} search states with 32 local conclusions per state, lower mean NLL is associated with higher local correctness, as shown in Figure~\ref{fig:preliminary-decay}. The protocol is described in Appendix~\ref{app:pilot-record}.

\begin{figure}[t]
    \centering
    \includegraphics[width=\columnwidth]{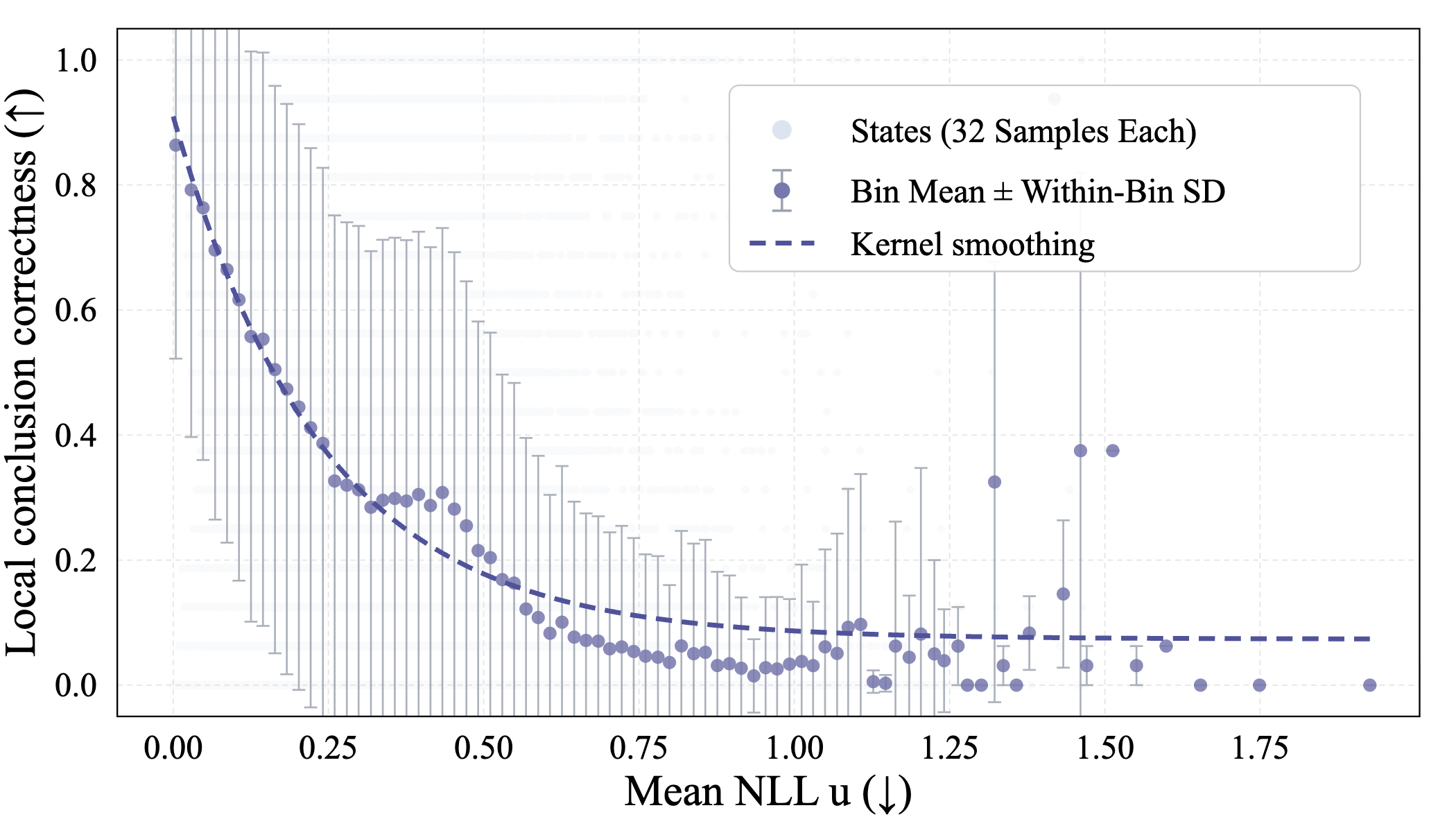}
    \caption{Lower mean NLL accompanies higher local correctness over 50,000 states, with 32 samples each. Faint points: state means; filled markers: bin means; bars: within-bin SD.}
    \Description{State-level mean local-conclusion correctness is plotted against mean NLL over 32 samples per state. Binned markers follow a decreasing trend; bars indicate one within-bin standard deviation.}
    \label{fig:preliminary-decay}
\end{figure}

\MainThreeBResults

\begin{figure*}[!t]
    \begin{minipage}[t]{\columnwidth}
        \vspace{0pt}
        \centering
        \includegraphics[width=\linewidth]{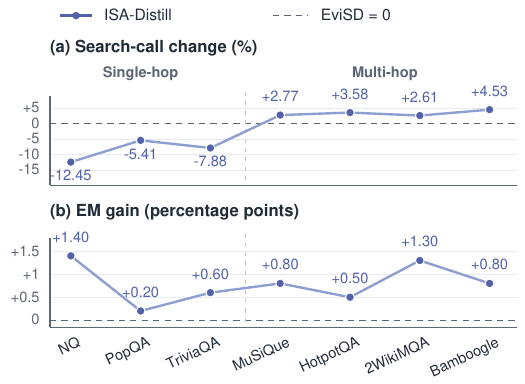}
        \caption{7B search allocation relative to EviSD (zero baseline). (a) Search-call change (\%), $100(C_{\mathrm{ISA}}/C_{\mathrm{EviSD}}-1)$. (b) EM gain in percentage points.}
        \Description{ISA-Distill reduces search calls on all three single-hop datasets. Calls rise on all four multi-hop datasets. Reported EM improves on all seven datasets.}
        \label{fig:search-allocation-7b}
    \end{minipage}\hfill
    \begin{minipage}[t]{\columnwidth}
        \vspace{0pt}
        \captionsetup{type=table}
        \centering
        \caption{Mean search calls per question with 7B models. $\Delta$ is relative to reproduced EviSD. Group and overall means weight dataset means by sample counts. Changes use underlying precision before display rounding.}
\label{tab:search-calls-7b}
\begingroup
\fontsize{8.5}{10.2}\selectfont
\setlength{\tabcolsep}{3pt}
\renewcommand{\arraystretch}{1.18}
\begin{tabularx}{\linewidth}{@{}>{\raggedright\arraybackslash}Xrrr@{}}
\toprule
\textbf{Dataset/group} & EviSD & ISA-Distill & $\Delta$ calls (\%) \\
\midrule
\rowcolor{baselinegroupblue}
\textbf{Single-hop avg.} & \textbf{2.14} & \textbf{1.99} & $\mathbf{-7.21}$ \\
NQ & 2.14 & 1.87 & $-12.45$ \\
PopQA & 2.20 & 2.08 & $-5.41$ \\
TriviaQA & 2.08 & 1.92 & $-7.88$ \\
\midrule
\rowcolor{baselinegroupblue}
\textbf{Multi-hop avg.} & \textbf{2.46} & \textbf{2.53} & $\mathbf{+2.94}$ \\
MuSiQue & 2.71 & 2.78 & $+2.77$ \\
HotpotQA & 2.32 & 2.40 & $+3.58$ \\
2WikiMQA & 2.49 & 2.56 & $+2.61$ \\
Bamboogle & 2.30 & 2.40 & $+4.53$ \\
\midrule
\textbf{Overall} & 2.28 & 2.22 & $-2.45$ \\
\bottomrule
\end{tabularx}
\endgroup

    \end{minipage}
\end{figure*}

We average the 32 NLL scores and map the result to a recoverability estimate $k\in[0,1]$:
\begin{equation}
u=\frac{1}{32}\sum_{r=1}^{32}U_{\pi}(\tilde y^{(r)}\mid P),\qquad k=f_{\mathrm{cal}}(u).
\label{eq:claim-nll}
\end{equation}
For each backbone, $f_{\mathrm{cal}}$ is fitted to local correctness by kernel smoothing before training and held fixed during policy optimization (Appendix~\ref{app:calibration-details}). We recompute $u$ and $k$ for each selected state--fact pair under the current rollout policy.

\subsection{Recoverability-Weighted Teacher Feedback}
\label{sec:teacher-routing}
The fact-use view $K$ receives the probe and the focal fact's value, with an instruction to use the fact and continue normally. The evidence-seeking view $S$ receives the same probe and an instruction to seek the corresponding evidence. These overlays enter teacher re-scoring of the original sampled action, as illustrated in Figure~\ref{fig:framework}.

When re-scoring search tokens, the original trajectory prefix excludes the current retrieval observation, preventing information returned by that search from entering the prefix. In the evidence-seeking view, masking applies only to the added overlay; facts already present in the history and the teacher's parametric knowledge remain available.

For an original sampled token $x$, let $q_0(x)$, $q_K(x)$, and $q_S(x)$ denote its probabilities under the original, fact-use, and evidence-seeking scoring contexts. Our self-distillation implementation uses the frozen rollout policy $\pi$ with the same causal prefix for all three scores. We bound each view's contrast with $q_0$ and mix them using $k$:
\begin{equation}
\begin{aligned}
z^K&=\tanh\!\left(\tau\log\frac{q_K(x)}{q_0(x)}\right),\\
z^S&=\tanh\!\left(\tau\log\frac{q_S(x)}{q_0(x)}\right),\\
c^{\mathrm{KS}}&=m\bigl[kz^K+(1-k)z^S\bigr].
\end{aligned}
\label{eq:teacher-contrast}
\end{equation}
The parameter $\tau>0$ controls contrast sensitivity, and $m\in\{0,1\}$ selects action tokens associated with the selected fact (Appendix~\ref{app:teacher-details}).

For search steps, the mask covers the opening search tag and query; for non-search intermediate steps, it covers the opening conclusion tag and the factual assertion associated with the focal fact.

Recoverability matters most when the two feedback views disagree. For fixed teacher scores, changing $k$ by $\Delta k$ changes the correction by $m\Delta k(z^K-z^S)$, leaving tokens with identical contrasts unaffected.

\subsection{Outcome-Anchored Policy Updates}
\label{sec:policy-learning}
Terminal correctness supplies reward $R\in\{0,1\}$ and group-relative advantage $A$. With $c=c^{\mathrm{KS}}\in[-1,1]$, we adjust this credit for each token~\cite{xie2026evisd}:
\begin{equation}
\widehat A=A+\lambda|A|\operatorname{sg}(c),\qquad 0\leq\lambda<1.
\label{eq:modulated-advantage}
\end{equation}
Here $\operatorname{sg}$ detaches the correction during optimization. Positive corrections strengthen positive outcome credit and soften negative credit. Because $|c|\leq1$ and $\lambda<1$, the correction preserves the advantage's sign. We use $\widehat A$ in clipped group relative policy optimization (GRPO), with a Kullback--Leibler (KL) penalty to a fixed reference policy~\cite{shao2024deepseekmath,schulman2017ppo}. Tokens outside the local mask retain their original outcome-based credit. The full objective is given in Appendix~\ref{app:optimization-details}.

Within each training batch, the frozen rollout policy supplies the student trajectories, local probe samples, and teacher token probabilities. Outcome rewards determine the group-relative advantages, and local feedback adjusts these advantages at the selected action tokens. At inference, only the student policy and retrieval tool are used.

\begin{figure}[t]
\centering
    \includegraphics[width=0.90\columnwidth]{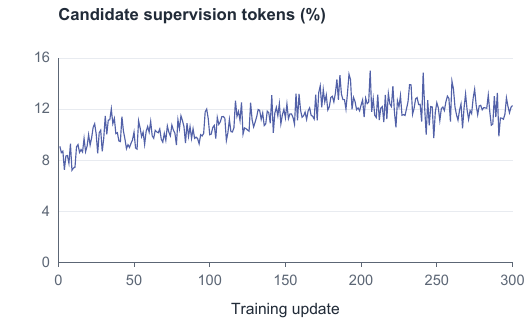}
    \caption{Candidate supervision content in Qwen-initialized 3B training. Query and non-search conclusion bodies occupy 11.32\% of generated tokens in saved prefixes. The line shows the ratio at each training update.}
    \Description{Combined candidate action-body token coverage rises from 9.43 percent over updates 1--50 to 12.07 percent over updates 251--300. Each update contributes its first 512 saved chunks.}
    \label{fig:candidate-supervision-coverage}
\end{figure}

\begin{table}[t]
\captionsetup{type=table,skip=6pt}
\centering
\caption{Routing ablations on Qwen2.5-3B-Instruct. Avg. is the macro-average EM (\%), weighting the seven datasets equally; calls are mean searches per evaluation trajectory.}
\label{tab:routing-ablations}
\begingroup
\fontsize{8.5}{10.2}\selectfont
\setlength{\tabcolsep}{3pt}
\renewcommand{\arraystretch}{1.0}
\begin{tabularx}{\columnwidth}{@{}>{\raggedright\arraybackslash}Xrr@{}}
\toprule
\textbf{Setting} & \textbf{Avg.} $\uparrow$ & \textbf{Calls} \\
\midrule
Base & 34.1 & 1.53 \\
K-only & 41.2 & 1.92 \\
S-only & 43.3 & 2.97 \\
Fixed mixture ($K/S=0.2/0.8$) & 41.3 & 2.13 \\
\midrule
\textbf{ISA-Distill (ours)} & 43.9 & 2.22 \\
\bottomrule
\end{tabularx}
\endgroup

\vspace{10pt}

\begingroup
\centering
\captionsetup{skip=4pt}
\caption{Macro-average EM across backbones. Avg. EM (\%) weights the seven benchmarks equally.}
\label{tab:backbone-comparison}
\fontsize{8.5}{10.2}\selectfont
\setlength{\tabcolsep}{2.8pt}
\renewcommand{\arraystretch}{1.0}
\begin{tabularx}{\columnwidth}{@{}Xr@{}}
\toprule
\textbf{Method} & \textbf{Avg. EM} $\uparrow$ \\
\midrule
\rowcolor{baselinegroupblue}
\multicolumn{2}{@{\hspace{3pt}}l@{}}{\fontsize{8}{9.2}\selectfont\color{black!80}\bfseries Qwen2.5-3B-Instruct} \\
EviSD & 42.6 \\
\textbf{ISA-Distill (ours)} & \textbf{43.9} \\
\addlinespace[1.5pt]
\rowcolor{baselinegroupblue}
\multicolumn{2}{@{\hspace{3pt}}l@{}}{\fontsize{8}{9.2}\selectfont\color{black!80}\bfseries Qwen2.5-7B-Instruct} \\
EviSD & 47.2 \\
\textbf{ISA-Distill (ours)} & \textbf{48.0} \\
\addlinespace[1.5pt]
\rowcolor{baselinegroupblue}
\multicolumn{2}{@{\hspace{3pt}}l@{}}{\fontsize{8}{9.2}\selectfont\color{black!80}\bfseries Llama-3B} \\
EviSD & 43.5 \\
\textbf{ISA-Distill (ours)} & \textbf{44.8} \\
\bottomrule
\end{tabularx}
\endgroup\space
\end{table}

\section{Experimental Setup}
\label{sec:experiments}
\label{sec:experimental-setup}

\subsection{Datasets and Evaluation Metrics}
\noindent\textbf{Datasets.}
We evaluate seven QA benchmarks. Natural Questions (NQ)~\cite{kwiatkowski2019natural} and HotpotQA~\cite{yang2018hotpotqa} provide in-domain evaluation; TriviaQA~\cite{joshi2017triviaqa}, PopQA~\cite{mallen2023trust}, 2WikiMultiHopQA (2WikiMQA)~\cite{ho2020constructing}, MuSiQue~\cite{trivedi2022musique}, and Bamboogle~\cite{press2023measuring} test cross-dataset transfer. We group NQ, TriviaQA, and PopQA as single-hop QA and the remaining four benchmarks as multi-hop QA. Evaluation splits and sample counts are listed in Appendix~\ref{app:dataset-splits}.

\noindent\textbf{Evaluation metrics.}
Answer accuracy is measured by exact match (EM), pooling correct predictions over all evaluated questions in each dataset. Avg. reports macro-average EM, weighting the seven datasets equally. Single-hop and multi-hop EM average the respective dataset scores. Search efficiency is measured by mean search calls per question, with task-group means weighted by dataset size.

\subsection{Baseline Definitions and Implementations}
\label{sec:baselines}

Baselines use Qwen2.5-3B-Instruct and Qwen2.5-7B-Instruct~\cite{qwen2024qwen25} and span outcome-based RL, local credit assignment, and self-distillation. We distinguish skill-conditioned and answer-conditioned RLSD.

\noindent\textbf{Outcome-based RL and reward shaping.}
\textit{Search-R1} learns interleaved reasoning and retrieval from final-answer correctness~\cite{jin2025searchr1}. \textit{AutoRefine} introduces explicit knowledge-refinement steps between searches and combines retrieval-specific rewards with answer-correctness rewards under GRPO~\cite{shi2025autorefine}.

\noindent\textbf{Turn- and step-level credit assignment.}
\textit{TIPS} shapes turn-level rewards using changes in a teacher's answer likelihood~\cite{xie2026tips}. \textit{PiCA} uses pivot steps to estimate trajectory-dependent search progress~\cite{liu2026pica}. \textit{GiGPO} estimates local relative advantages by grouping actions from repeated environment states~\cite{feng2025gigpo}.

\noindent\textbf{Rollout-conditioned self-distillation.}
\textit{SD-Search} constructs hindsight context from a rollout group's search queries and correctness labels, then aligns search-query token distributions with a hindsight-conditioned self-teacher~\cite{ma2026sdsearch}.

\noindent\textbf{Skill-conditioned self-distillation.}
\textit{RLSD}$_{\mathrm{Skill}}$ uses skill-conditioned teacher inputs to modulate outcome-based token advantages~\cite{yang2026rlsd}. \textit{Skill-SD} conditions its teacher on trajectory-derived skills and combines GRPO with reverse-KL distillation~\cite{wang2026skillsd}. \textit{SDAR} incorporates a gated skill-conditioned distillation objective into RL~\cite{lu2026sdar}.

\noindent\textbf{Answer-conditioned self-distillation.}
\textit{RLSD}$_{\mathrm{Ans}}$ supplies gold answers to the teacher for advantage modulation~\cite{yang2026rlsd}.

\noindent\textbf{Evidence-conditioned self-distillation.}
\textit{EviSD} conditions search-action feedback on instance-level supporting evidence and answer-action feedback on gold answers. It converts privileged token-level contrasts into bounded adjustments to outcome-based advantages~\cite{xie2026evisd}.

\noindent\textbf{Additional details.}
Appendix~\ref{app:experimental-details} provides dataset splits, training settings, and diagnostic protocols.

\begin{figure*}[t]
    \centering
    \includegraphics[width=0.92\textwidth]{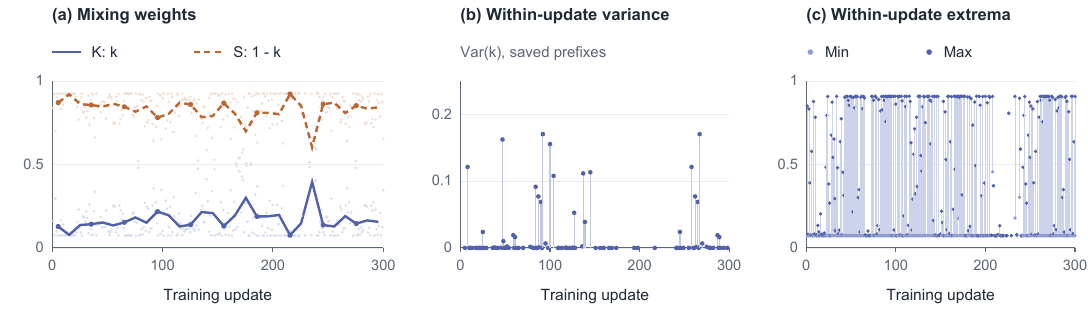}
    \caption{Recoverability-weighted feedback. (a) Probe-weighted 10-update means of $k$ and $1-k$; faint points are per-update means. (b) Within-update variance from saved prefixes, available for 96 updates with at least two probes. (c) Per-update minima and maxima of $k$.}
    \Description{Three panels share the training-update axis. The average evidence-seeking weight exceeds the fact-use weight throughout the pooled training windows. Saved-prefix within-update variances range from nearly zero to 0.171. The displayed extrema contain both k below 0.1 and k above 0.9 in 91 of the 286 updates with probes. Variance and extrema use different observation scopes, as specified in the caption.}
    \label{fig:recovery-weight-dynamics}
\end{figure*}

\begin{figure}[!t]
\centering
        \includegraphics[width=\columnwidth]{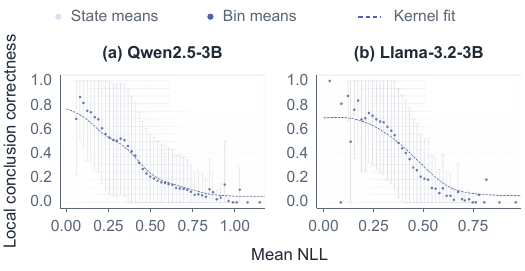}
        \caption{Local-conclusion correctness decreases nonlinearly with mean NLL for (a) Qwen2.5-3B and (b) Llama-3.2-3B. Each search state uses 32 conclusions; bars indicate within-bin SD.}
        \Description{Two panels plot state-level mean local-conclusion correctness against mean NLL. Qwen2.5-3B declines over a broader NLL range. Llama-3.2-3B has a relatively flat fitted curve at low NLL followed by a more pronounced middle-range decline. Both approach low local correctness at high NLL. Faint points denote state means, darker points denote bin means, error bars indicate within-bin standard deviation, and dashed curves denote kernel fits.}
        \label{fig:backbone-uncertainty}

\vspace{8pt}

\centering
    \includegraphics[width=0.95\columnwidth]{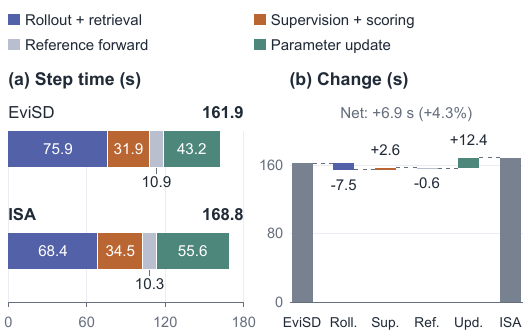}
    \caption{Mean training-step time. (a) Stacked component times and their totals. (b) Additive changes from EviSD to ISA-Distill (ISA); endpoint bars show total time. Supervision includes preparation and scoring.}
    \Description{EviSD takes 161.9 seconds per training step: 75.9 for rollout generation and retrieval, 31.9 for supervision preparation and scoring, 10.9 for the reference forward pass, and 43.2 for parameter updates. ISA-Distill takes 168.8 seconds: 68.4, 34.5, 10.3, and 55.6 seconds for the same stages. The changes are minus 7.5, plus 2.6, minus 0.6, and plus 12.4 seconds, respectively, yielding a net increase of 6.9 seconds or 4.3 percent.}
    \label{fig:training-cost}
\end{figure}

\begin{figure*}[!t]
\centering
\includegraphics[width=0.85\textwidth]{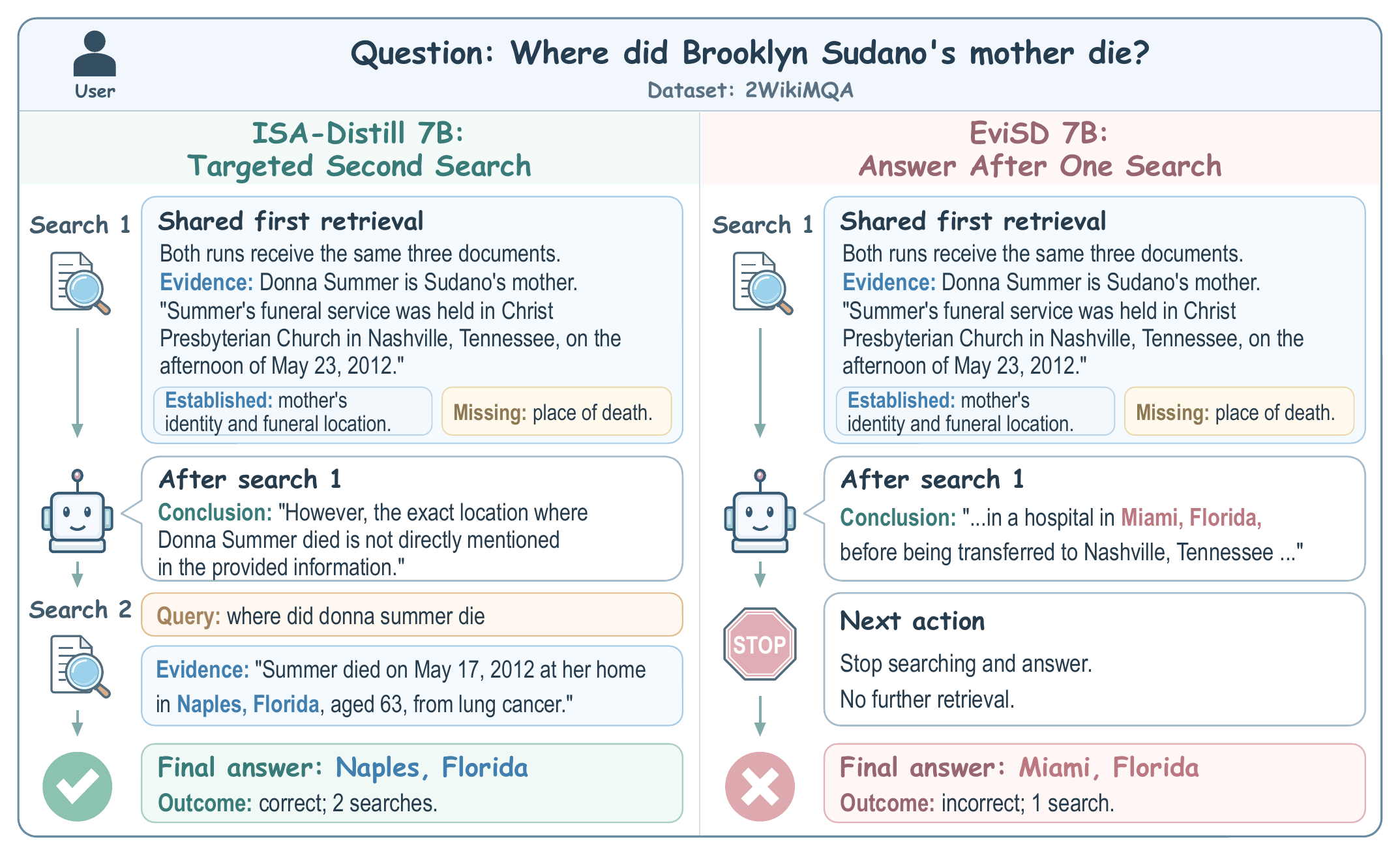}
\caption{ISA-Distill retrieves the missing fact; EviSD gives an unsupported answer from the same initial evidence.}
\Description{Trajectory comparison for the question Where did Brooklyn Sudano's mother die. Identical first-search documents establish that Donna Summer is the mother and her funeral was in Nashville, but do not state where she died. ISA-Distill identifies the missing location, searches where did donna summer die, retrieves evidence for Naples, Florida, and answers correctly after two searches. EviSD claims a Miami hospital and transfer to Nashville in its conclusion, performs no further retrieval, and answers Miami incorrectly after one search.}
\label{fig:isa-case-study}
\end{figure*}

\section{Results and Analysis}
\label{sec:results}

\subsection{Main Results}
\label{sec:main-results}
\label{sec:search-behavior}

\noindent\textbf{Accuracy gains across model scales.} ISA-Distill achieves the highest macro-average EM among the evaluated methods at both 3B and 7B (Tables~\ref{tab:main-7b} and~\ref{tab:main-3b}). At 3B, the improvement over EviSD is concentrated in multi-hop QA, where mean EM increases by 2.1 points while single-hop accuracy remains nearly unchanged. At 7B, gains over EviSD extend across all seven benchmarks, covering both in-domain and cross-dataset evaluation. These gains support the generalizability of ISA-Distill across model scales and datasets.

\noindent\textbf{More search on multi-hop tasks, less on single-hop tasks.} Compared with reproduced EviSD, the 7B model makes 7.21\% fewer searches on single-hop tasks and 2.94\% more on multi-hop tasks, while reducing calls overall by 2.45\%, as shown in Figure~\ref{fig:search-allocation-7b} and Table~\ref{tab:search-calls-7b}. Higher accuracy thus accompanies different changes in search frequency across task groups.

\subsection{Training Feedback Analysis}
\label{sec:training-feedback}
\noindent\textbf{Sparse supervision of key action tokens supports strong performance.} ISA-Distill achieves 43.9 macro EM at 3B (Table~\ref{tab:main-3b}), while candidate spans for supervising key action tokens account for only 11.32\% of generated tokens in saved prefixes. Figure~\ref{fig:candidate-supervision-coverage} shows that this coverage increases from 9.43\% in the first 50 updates to 12.07\% in the last 50.

\noindent\textbf{Average weights favor evidence seeking, while local weights can strongly favor either view.} As shown in Figure~\ref{fig:recovery-weight-dynamics}(a), the 10-update mean weights consistently favor $S$, with overall means of 0.163 for $K$ and 0.837 for $S$. Figure~\ref{fig:recovery-weight-dynamics}(b) shows that within-update variance is usually low but occasionally reaches 0.171 in saved prefixes. As shown in Figure~\ref{fig:recovery-weight-dynamics}(c), both $k<0.1$ and $k>0.9$ occur in 91 updates: different state--fact pairs within the same update can strongly favor opposite teacher views.

\noindent\textbf{Local weighting can change the policy gradient even when the mean mixture is stable.} Holding a sampled batch and its detached supervision fixed, we compare local weights with a shared weight $\bar k$, the batch mean of the recoverability estimates. For token $j$ in a matched span, let $k_j$ and $A_j$ denote its state--fact weight and rollout advantage. Equations~\ref{eq:teacher-contrast}--\ref{eq:modulated-advantage} give the local-minus-shared advantage difference
\begin{equation}
\Delta\widehat A_j
=\lambda|A_j|m_j(k_j-\bar k)(z_j^K-z_j^S).
\label{eq:local-mean-advantage}
\end{equation}
Evaluate both policy gradients at $\theta=\theta_{\mathrm{old}}$, where the importance ratio is one and clipping is inactive. The shared KL term cancels, so the objective in Appendix~\ref{app:optimization-details} yields
\begin{equation}
\Delta g
=\sum_j w_j\Delta\widehat A_j
\left.\nabla_\theta\log\pi_\theta(x_j\mid\xi_j)
\right|_{\theta=\theta_{\mathrm{old}}},
\label{eq:local-mean-gradient}
\end{equation}
where $w_j$ are the averaging weights in the empirical batch objective, $\xi_j$ is the student-visible prefix, and $\Delta g$ is the local-minus-shared objective gradient. Local weights act on teacher contrasts and token gradients before aggregation, so the batch gradient depends on their state-specific assignment.

\subsection{Ablation Study}
\label{sec:routing-eval}
These ablations support the importance of recoverability-guided weighting in balancing fact-use and evidence-seeking feedback. Neither single-view supervision nor a fixed mixture matches the answer accuracy achieved by adapting the weights to the student's current information state.

\noindent\textbf{Ablation of feedback views.} Recoverability-based weighting improves accuracy with fewer searches than S-only feedback. ISA-Distill reaches 43.9 macro EM with 2.22 searches per question, versus 43.3 EM and 2.97 searches for S-only (Table~\ref{tab:routing-ablations}): a 0.6-point gain with approximately 25.3\% fewer searches. S-only nearly exhausts the three-call limit.

\noindent\textbf{Ablation of adaptive weighting.} A fixed mixture reduces search frequency at the cost of accuracy. We set $K=0.2$ and $S=0.8$ to approximate the mean training weights in Section~\ref{sec:training-feedback}. This reduces mean calls to 2.13 but lowers macro EM to 41.3. The gap to ISA-Distill supports the practical value of local weighting, consistent with the analysis in Section~\ref{sec:training-feedback} that local weights can change the policy gradient even when the mean mixture is stable.

\subsection{Cross-Backbone Analysis}
\label{sec:cross-backbone}
\label{sec:backbone-uncertainty}

\noindent\textbf{Accuracy gains extend across the evaluated backbones.} ISA-Distill improves macro-average EM over EviSD by 1.3 points on Qwen2.5-3B, 0.8 on Qwen2.5-7B, and 1.3 on Llama-3B~\cite{meta2024llama32} (Table~\ref{tab:backbone-comparison}). The gains on both Qwen and Llama support the generalizability of ISA-Distill across backbone families.

\noindent\textbf{The relationship between NLL and local correctness varies across backbones.} As shown in Figure~\ref{fig:backbone-uncertainty}, under the same 32-conclusion protocol, local correctness decreases with NLL for Qwen2.5-3B-Instruct and Llama-3.2-3B-Instruct, but the curves differ in shape and range, motivating separate calibration for each backbone.

\subsection{Training Efficiency Analysis}
\label{sec:training-cost}
ISA-Distill adds 4.3\% to measured training-step time. Mean step time increases from 161.9\,s for EviSD to 168.8\,s for ISA-Distill, as shown in Figure~\ref{fig:training-cost}. Relative to EviSD, ISA-Distill introduces additional supervision computation by constructing complementary fact-use and evidence-seeking teacher views, re-scoring sampled student actions under both views, and estimating recoverability to weight their feedback. The measured net overhead includes changes in both supervision and policy optimization costs, partly offset by lower rollout and retrieval time. These additional computations are confined to training; inference uses only the student and retrieval tool.

\subsection{Case Study}
\label{sec:case-study}
Figure~\ref{fig:isa-case-study} compares two 7B trajectories with identical initial evidence (Appendix~\ref{app:case-records}), highlighting how ISA-Distill acquires missing evidence before answering. The documents establish Donna Summer as Brooklyn Sudano's mother and locate her funeral in Nashville, but do not state where she died. ISA-Distill notes that the death location is missing and retrieves a passage identifying Naples as the place of death through a targeted second search. EviSD instead claims without evidence that Summer died in a Miami hospital and answers incorrectly after one search. The second search builds on the established identity to obtain evidence for the remaining part of the question.

\section{Conclusion}
We identify information-state mismatch in on-policy distillation: privileged knowledge can lead the teacher to undervalue searches that help the student acquire missing evidence. ISA-Distill addresses this mismatch by using factual recoverability to balance fact-use and evidence-seeking feedback. In controlled experiments, ISA-Distill improves the accuracy--search trade-off and reallocates retrieval toward multi-hop tasks. These findings suggest that teacher feedback for search agents should reflect the student's information needs at each step. Such feedback should help the student use available knowledge and acquire evidence for facts it cannot reliably recover.

\clearpage
\balance

\clearpage
\nobalance
\appendix
\section{Method Implementation}
\label{app:method-details}
We use $t$ for steps, $\ell$ for tokens, and $i$ for rollouts; $\pi_{\mathrm{old}}$ denotes the frozen rollout policy.

\subsection{Interaction Format and State}
\label{sec:information-access}
\paragraph{Interaction prompt.}
Following prior work~\cite{wu2026hiprag}, we use a structured interaction format (Table~\ref{tab:interaction-prompt}). We denote reasoning, query, environment observation, and conclusion by $r$, $w$, $o$, and $y$, respectively.

\begin{table}[!t]
\centering
\caption{Task instructions and step formats from the structured interaction prompt.}
\label{tab:interaction-prompt}
\begin{minipage}[t]{\linewidth}
\vspace{0pt}
\setlength{\fboxsep}{4pt}
\colorbox{black}{%
    \parbox{\dimexpr\linewidth-2\fboxsep\relax}{%
        \color{white}\fontsize{8.5}{10}\selectfont\bfseries Interaction Prompt: Task and Step Formats
    }%
}\\[-0.4pt]
\hspace*{\dimexpr\fboxsep+\fboxrule\relax}%
\begin{minipage}{\dimexpr\linewidth-2\fboxsep-2\fboxrule\relax}
\begin{lstlisting}[
    numbers=none,
    frame=single,
    rulecolor=\color{black},
    backgroundcolor=\color{gray!10},
    basicstyle=\fontsize{8}{9.6}\selectfont\ttfamily,
    columns=fullflexible,
    keepspaces=true,
    breakindent=0pt,
    xleftmargin=0pt,
    framesep=4pt,
    framexleftmargin=0pt,
    framexrightmargin=0pt,
    framextopmargin=4pt,
    framexbottommargin=4pt,
    aboveskip=0pt,
    belowskip=0pt,
    breaklines=true,
    breakatwhitespace=true,
    showstringspaces=false
]
Solve the question using factual steps and, when needed, the search tool. The complete interaction has exactly one <think> block containing one or more <step> blocks, followed by exactly one <answer> block outside <think>.

Each step addresses one local information need. First use <reasoning> for a brief assessment of what is known and what is missing. End that step with a <conclusion> stating its result. Search only when evidence is needed; a step without search omits BOTH search and context tags, including empty placeholders.

Search-step layout:
<step>
<reasoning>brief assessment of the information need</reasoning>
<search>a specific query</search>
<context>results inserted only by the environment</context>
<conclusion>what the returned evidence supports, or what remains unresolved</conclusion>
</step>

Non-search-step layout:
<step>
<reasoning>brief assessment using current knowledge and visible evidence</reasoning>
<conclusion>a local factual conclusion</conclusion>
</step>

After the last completed step, emit:
</think>
<answer>only the short final answer to the original question</answer>
\end{lstlisting}
\end{minipage}
\end{minipage}
\end{table}

\paragraph{Pre-step state.}
The rollout saves $s_t^-$ from Equation~\ref{eq:preaction-state} before each step. Each retrieved observation is available when generating the corresponding conclusion and subsequent steps. At $b_t=0$, a shared grammar mask removes search actions from rollout and scoring; malformed queries are not executed.

\subsection{Privileged Information Selection}
\label{app:focus-details}
\label{sec:focus-selection}
Given the saved pre-step state $s_t^-$ and sampled action, the selector chooses an action-relevant focal fact from the dataset's human-annotated supporting evidence and gold answers. It formulates a value-masked probe that asks for this fact using entities and relations from the visible context.

\subsection{Recoverability Estimation}
\label{app:calibration-details}
\paragraph{Probe sampling and scoring.}
The input consists of $s_t^-$ followed by a value-masked probe, with the current step's reasoning and query excluded. The probe omits the target value and its reference-only aliases (Table~\ref{tab:recovery-probe-prompt}). The frozen rollout policy $\pi_{\mathrm{old}}$ independently generates 32 local conclusions for this probe without further retrieval. We compute each conclusion's mean token NLL, excluding padding, and average the 32 scores to obtain $u$ in Equation~\ref{eq:claim-nll}.

\begin{table}[!t]
\centering
\caption{Value-masked probe suffix appended to the saved pre-step state. The selector supplies the value-masked \texttt{\{probe\}}; the student generates 32 local conclusions from this prompt.}
\label{tab:recovery-probe-prompt}
\begin{minipage}[t]{\linewidth}
\vspace{0pt}
\setlength{\fboxsep}{4pt}
\colorbox{black}{%
    \parbox{\dimexpr\linewidth-2\fboxsep\relax}{%
        \color{white}\fontsize{8.5}{10}\selectfont\bfseries Value-Masked Probe
    }%
}\\[-0.4pt]
\hspace*{\dimexpr\fboxsep+\fboxrule\relax}%
\begin{minipage}{\dimexpr\linewidth-2\fboxsep-2\fboxrule\relax}
\begin{lstlisting}[
    numbers=none,
    frame=single,
    rulecolor=\color{black},
    backgroundcolor=\color{gray!10},
    basicstyle=\fontsize{8}{9.6}\selectfont\ttfamily,
    columns=fullflexible,
    keepspaces=true,
    breakindent=0pt,
    xleftmargin=0pt,
    framesep=4pt,
    framexleftmargin=0pt,
    framexrightmargin=0pt,
    framextopmargin=4pt,
    framexbottommargin=4pt,
    aboveskip=0pt,
    belowskip=0pt,
    breaklines=true,
    breakatwhitespace=true,
    showstringspaces=false
]
[Diagnostic factual question]
Pause the original task. Do not search. Answer the following question using the preceding visible state and your knowledge. Output only the short answer, no XML or explanation.
{probe}
Answer:
\end{lstlisting}
\end{minipage}
\end{minipage}
\end{table}

\paragraph{Fixed calibration map.}
Given calibration-bin state counts $n_j$, mean NLLs $\Phi_j$, and mean local correctness $\Psi_j$ (Appendix~\ref{app:recovery-calibration-protocol}), define the count-weighted Gaussian smoother:
\begin{equation}
\begin{aligned}
k=f_{\mathrm{cal}}(u)
&=\frac{\sum_{j\in\mathcal J_{50}}n_j\Psi_j W_h(u,\Phi_j)}
{\sum_{j\in\mathcal J_{50}}n_j W_h(u,\Phi_j)},\\
W_h(u,\Phi)
&=e^{-\frac{(u-\Phi)^2}{2h^2}}
+e^{-\frac{(u+\Phi)^2}{2h^2}}.
\end{aligned}
\label{eq:claim-calibration}
\end{equation}
Here $\mathcal J_{50}=\{j:n_j\geq50\}$ and $h=0.055$. The second term reflects observations at the nonnegative NLL boundary. For nonempty $\mathcal J_{50}$, positive normalized weights keep $k\in[0,1]$; otherwise the estimate is unavailable.

We recompute the mean NLL $u$ and recoverability estimate $k$ for each selected state--fact pair under the current rollout snapshot (Appendix~\ref{app:recovery-calibration-protocol}).

\subsection{Teacher Contexts and Token Scoring}
\label{app:teacher-details}
The self-distillation teacher $\pi_T=\pi_{\mathrm{old}}$ shares the student's tokenizer and scoring support.

\paragraph{Teacher overlays.}
The fact-use view $D_t^K$ supplies the selector's probe and its canonical answer; the evidence-seeking view $D_t^S$ supplies the same probe without the answer and requests external evidence (Table~\ref{tab:fact-use-instruction}).

\begin{table}[!t]
\centering
\caption{Teacher overlays prepended to the original scoring context. Both views receive the selector's \texttt{\{probe\}}; K also receives the canonical \texttt{\{target\}}.}
\label{tab:fact-use-instruction}
\begin{minipage}[t]{\linewidth}
\vspace{0pt}
\setlength{\fboxsep}{4pt}
\colorbox{black}{%
    \parbox{\dimexpr\linewidth-2\fboxsep\relax}{%
        \color{white}\fontsize{8.5}{10}\selectfont\bfseries Teacher Overlays: Fact Use (K) and Evidence Seeking (S)
    }%
}\\[-0.4pt]
\hspace*{\dimexpr\fboxsep+\fboxrule\relax}%
\begin{minipage}{\dimexpr\linewidth-2\fboxsep-2\fboxrule\relax}
\begin{lstlisting}[
    numbers=none,
    frame=single,
    rulecolor=\color{black},
    backgroundcolor=\color{gray!10},
    basicstyle=\fontsize{8}{9.6}\selectfont\ttfamily,
    columns=fullflexible,
    keepspaces=true,
    breakindent=0pt,
    xleftmargin=0pt,
    framesep=4pt,
    framexleftmargin=0pt,
    framexrightmargin=0pt,
    framextopmargin=4pt,
    framexbottommargin=4pt,
    aboveskip=0pt,
    belowskip=0pt,
    breaklines=true,
    breakatwhitespace=true,
    showstringspaces=false
]
[Training-only factual information]
Question: {probe}
Correct factual answer: {target}
Use this fact to continue the original task normally.

[Training-only information need]
{probe}
Seek external evidence to resolve this information need. No answer is supplied. Continue from the original context.
\end{lstlisting}
\end{minipage}
\end{minipage}
\end{table}

The reference content in each overlay is restricted to the selected focal fact: $D_t^K$ adds its value and relation binding, while $D_t^S$ masks the value and its reference-only aliases. Masking applies to the added overlay; existing target mentions and the teacher's parametric knowledge remain available.

\paragraph{Matched scoring contexts.}
For $B\in\{0,K,S\}$ and $D_t^0=\varnothing$, the scoring contexts share the question, history, completed observations, budget, truncation, grammar mask, and original causal prefix:
\begin{equation}
q_B(\cdot\mid s_t^-,x_{t,<\ell})
=\pi_T(\cdot\mid D_t^B\oplus s_t^-,x_{t,<\ell}).
\label{eq:contexts}
\end{equation}
A shared context builder reserves space for the maximum overlay length before rollout. The causal prefix $x_{t,<\ell}$ includes current-step reasoning when applicable; search-token re-scoring ends before the current observation. Both views preserve the original action support, including non-search actions under the evidence-seeking instruction. Overlays enter only the scoring contexts.

\paragraph{Correction spans.}
For search steps, the local mask selects the \texttt{<search>} opening tag and query. For non-search intermediate steps, it selects the \texttt{<conclusion>} opening tag and the factual assertion about the selected focal fact. Exact byte offsets are mapped to tokenizer spans; ambiguous or cross-boundary mappings are rejected. Local correction applies only to these spans. Missing or invalid focal facts, spans, probes, overlays, or recoverability estimates give $c^{\mathrm{KS}}=0$.

\subsection{Policy Optimization}
\label{app:optimization-details}
\label{app:training-details}
\paragraph{Outcome credit.}
Terminal correctness is verified against benchmark answer aliases; malformed output or truncation without a valid final answer receives zero, while earlier valid search corrections remain fixed. For a group of $G$ rollouts with rewards $R_i=R_{\mathrm{task}}(\tau_i)\in\{0,1\}$, the outcome advantage is
\begin{equation}
A_i=\frac{R_i-\overline R}
{\sqrt{\frac{1}{G}\sum_{r=1}^G(R_r-\overline R)^2}+\epsilon_A},
\label{eq:task-advantage}
\end{equation}
where $\overline R=G^{-1}\sum_{r=1}^G R_r$ and $\epsilon_A>0$. All-correct and all-failed groups have $A=\widehat A=0$, leaving only the KL update.

\paragraph{Actor objective.}
For a student-generated rollout token $x_{in}$ with original visible prefix $\xi_{in}$, let $\rho_{in}(\theta)=\pi_\theta(x_{in}\mid\xi_{in})/\pi_{\mathrm{old}}(x_{in}\mid\xi_{in})$. Let $\mathcal T_i$ denote the generated-token indices in rollout $i$. We maximize the following objective over $\theta$~\cite{shao2024deepseekmath,schulman2017ppo}:
\begin{equation}
\begin{aligned}
\mathcal J(\theta)=\mathbb E\Biggl[
\frac{1}{G}\sum_{i=1}^G\frac{1}{|\mathcal T_i|}
\sum_{n\in\mathcal T_i}
\Bigl(&\min\{\rho_{in}\widehat A_{in},\\[-2pt]
&\operatorname{clip}(\rho_{in},1-\varepsilon,1+\varepsilon)
\widehat A_{in}\}\\
&-\beta_{\mathrm{KL}}\operatorname{KL}_{in}(\theta)\Bigr)
\Biggr],
\end{aligned}
\label{eq:single-grpo}
\end{equation}
where $\operatorname{KL}_{in}$ is conditional KL to a fixed reference policy or its standard estimator, and the actor loss is $-\mathcal J$.

The rollout snapshot, teacher, and detached supervision remain fixed across each batch's optimization epochs; the next batch refreshes the snapshot.

\section{Experimental Details}
\label{app:experimental-details}
\subsection{Dataset Splits and Sample Counts}
\label{app:dataset-splits}
Evaluation covers 51,713 questions across seven QA benchmarks (Table~\ref{tab:evaluation-splits}).

\begin{table}[!t]
\centering
\caption{Evaluation splits and sample counts for the seven QA benchmarks.}
\label{tab:evaluation-splits}
\small
\renewcommand{\arraystretch}{1.08}
\begin{tabular*}{\columnwidth}{@{\extracolsep{\fill}}lcr@{}}
\toprule
\textbf{Dataset} & \textbf{Split used} & \textbf{Samples} \\
\midrule
\rowcolor{baselinegroupblue}
\multicolumn{3}{l}{\textbf{Single-Hop QA}} \\
NQ~\cite{kwiatkowski2019natural} & test & 3,610 \\
TriviaQA~\cite{joshi2017triviaqa} & test & 11,313 \\
PopQA~\cite{mallen2023trust} & test & 14,267 \\
\midrule
\rowcolor{baselinegroupblue}
\multicolumn{3}{l}{\textbf{Multi-Hop QA}} \\
HotpotQA~\cite{yang2018hotpotqa} & dev & 7,405 \\
2WikiMultiHopQA~\cite{ho2020constructing} & dev & 12,576 \\
MuSiQue~\cite{trivedi2022musique} & dev & 2,417 \\
Bamboogle~\cite{press2023measuring} & test & 125 \\
\midrule
\multicolumn{2}{l}{\textbf{Total}} & \textbf{51,713} \\
\bottomrule
\end{tabular*}
\end{table}

\subsection{Training and Decoding Hyperparameters}
\label{app:hyperparameters}
Batch size counts input questions, while policy mini-batches count generated chunks (Table~\ref{tab:training-hyperparameters}). Dynamic batching enforces the actor's token limit per GPU.

\begin{table}[!t]
\centering
\caption{Principal training and decoding hyperparameters.}
\label{tab:training-hyperparameters}
\begingroup
\small
\setlength{\tabcolsep}{3pt}
\renewcommand{\arraystretch}{1.08}
\begin{tabularx}{\columnwidth}{@{}Xr@{}}
\toprule
\textbf{Hyperparameter} & \textbf{Value} \\
\midrule
Training updates & 300 \\
GPUs & 4 \\
Train batch size (questions) & 128 \\
Rollouts per question $G$ & 8 \\
Learning rate & $1\times10^{-6}$ \\
Parameter updates & Full-parameter training \\
Policy mini-batch size (chunks) & 256 \\
Actor token budget per GPU & 8,192 \\
Policy epochs per update & 1 \\
PPO clipping parameter $\varepsilon$ & 0.2 \\
KL coefficient $\beta_{\mathrm{KL}}$ & 0.001 \\
\midrule
Rollout temperature & 1.0 \\
Maximum search calls & 3 \\
Retriever top-$k$ & 3 \\
State context limit (tokens) & 4,096 \\
Generation chunk limit (tokens) & 512 \\
Trajectory limit (tokens) & 8,192 \\
\midrule
Local correction strength $\lambda$ & 0.2 \\
Contrast scale $\tau$ & 5.0 \\
Samples per value-masked probe & 32 \\
\bottomrule
\end{tabularx}
\endgroup
\end{table}

\subsection{Calibration Data and Map Construction}
\label{app:recovery-calibration-protocol}
\paragraph{Data and sampling.}
Calibration uses the local-state sampling and assessment protocol in Appendix~\ref{app:pilot-record}, with the same input format used for recoverability estimation during training: the saved pre-step state followed by a value-masked probe. Each backbone is frozen during sampling; reference answers enter assessment only.

\paragraph{Aggregation.}
Let $U_{i,r}$ be the mean token NLL of conclusion $r$ at state $i$, excluding padding, and $A_{i,r}$ its binary local correctness under the assessment in Appendix~\ref{app:pilot-record}. State-level means are
\begin{equation}
\overline U_i=\frac{1}{32}\sum_{r=1}^{32}U_{i,r},
\qquad
\overline A_i=\frac{1}{32}\sum_{r=1}^{32}A_{i,r}.
\label{eq:pilot-state-means}
\end{equation}

\paragraph{Map construction.}
We partition the states' $\overline U_i$ range into 100 equal-width bins. For each nonempty bin $B_j$ containing $n_j$ states, the centroid is
\begin{equation}
\Phi_j=\frac{1}{n_j}\sum_{i\in B_j}\overline U_i,
\qquad
\Psi_j=\frac{1}{n_j}\sum_{i\in B_j}\overline A_i.
\label{eq:pilot-bin-centroids}
\end{equation}
Equation~\ref{eq:claim-calibration} constructs one calibration map per backbone from these summaries. Bin summaries and bandwidth remain fixed during policy optimization.

\paragraph{Training-time use.}
For each value-masked probe, the current rollout policy generates 32 local conclusions. We average their mean token NLLs and apply the fixed calibration map to obtain the recoverability estimate $k$, which weights the two teacher views (Appendix~\ref{app:calibration-details}).

\subsection{NLL and Local Correctness: Curve Construction}
\label{app:pilot-record}
\paragraph{State sampling.}
Using a fixed seed, we sample 50,000 questions from existing Qwen2.5-7B-Instruct trajectories and select one valid search state per question. Each input pairs the saved pre-step state with a value-masked probe (Appendix~\ref{app:calibration-details}), excluding the current step's reasoning, query, retrieval evidence, and original conclusion, as well as reference answers.

\paragraph{Generation and assessment.}
We independently sample 32 local conclusions per state and compute each conclusion's mean token NLL, excluding artificially added padding. Qwen3.7-Flash~\cite{qwen2026flash} checks the local reference against the dataset's human annotations, then judges whether each conclusion fully answers the current local objective.

\paragraph{Aggregation.}
We bin the state means in Equation~\ref{eq:pilot-state-means} by NLL and report bin means with within-bin standard deviations. We apply the same protocol to Qwen2.5-3B-Instruct and Llama-3.2-3B-Instruct, with results shown in Figure~\ref{fig:backbone-uncertainty}.

\subsection{Case Study}
\label{app:case-records}
The comparison uses item \texttt{2wikimultihopqa:\allowbreak eval\_43819} from 2WikiMultiHopQA~\cite{ho2020constructing}, with accepted answers ``Naples'' and ``Naples, Florida''. The ISA-Distill trajectory is from the 7B model; the comparator is EviSD 7B. Their first queries differ only in capitalization and return identical documents in the same order.

\subsection{Paired Information-State Diagnostic}
\label{app:diagnostic-protocols}
\label{app:paired-diagnostic}

\paragraph{Sample and factual support.}
The diagnostic uses 698 state--search-action pairs from 228 HotpotQA training questions. Eligible pairs have a reference-supported focal fact, an associated query, and identifiable support in the actual retrieval. Facts proposed by a local 7B model were checked and, where needed, corrected by an AI assistant against reference sentences and document titles.

\paragraph{Teacher scores.}
Both teacher and student use the retained Qwen2.5-3B checkpoint at update 300, with pre-step history, current reasoning, and original search tokens held fixed. Let $\ell_i(c)$ be the mean teacher log-probability of pair $i$'s search-opening and query tokens under added context $c$. We measure
\[
\Delta_{\mathrm{priv},i}=\ell_i(f_i)-\ell_i(\varnothing),
\]
where $f_i$ supplies the focal fact. All conditions use identical behavioral instructions. Two fact wordings are each paired with non-target factual text of exactly the same tokenizer length. Scores use FP32 with TF32 disabled. The diagnostic changes explicit fact access; the method's fact-use and evidence-seeking views additionally change behavioral instructions.

\paragraph{Evidence utility.}
We continue the frozen student with either the actual retrieval or the same retrieval with the designated focal-support spans masked. Content outside those spans is unchanged. Each condition uses 16 draws at temperature 1 with matched seeds and identical remaining search, step, and trajectory-token budgets; masking does not refund tokens. Later queries may differ under these budgets. With final-answer correctness $R_{ij}\in\{0,1\}$, observed utility is
\[
\widehat V_i=\frac{1}{16}\sum_{j=1}^{16}
\bigl(R_{ij}^{\mathrm{full}}-R_{ij}^{\mathrm{masked}}\bigr).
\]
Invalid output formats count as incorrect.

\paragraph{Controls and uncertainty.}
Against length-matched non-target text, 258 of 396 positive-utility pairs receive lower scores (65.2\%; 95\% CI: 58.8--71.3\%). Requiring a decrease exceeding 0.01 nats/token under both fact wordings and both baselines retains 171 pairs (43.2\%; 95\% CI: 37.2--49.2\%). Among the 293 pairs with positive utility in both eight-draw halves, 183 receive lower scores (62.5\%; 95\% CI: 54.7--70.0\%). Intervals are percentiles from 5,000 bootstrap resamples of connected question/reference-title groups (seed 20261018), which coincide with question groups here. They condition on the observed utility classification.


\begin{thebibliography}{37}


\ifx \showCODEN    \undefined \def \showCODEN     #1{\unskip}     \fi
\ifx \showISBNx    \undefined \def \showISBNx     #1{\unskip}     \fi
\ifx \showISBNxiii \undefined \def \showISBNxiii  #1{\unskip}     \fi
\ifx \showISSN     \undefined \def \showISSN      #1{\unskip}     \fi
\ifx \showLCCN     \undefined \def \showLCCN      #1{\unskip}     \fi
\ifx \shownote     \undefined \def \shownote      #1{#1}          \fi
\ifx \showarticletitle \undefined \def \showarticletitle #1{#1}   \fi
\ifx \showURL      \undefined \def \showURL       {\relax}        \fi
\providecommand\bibfield[2]{#2}
\providecommand\bibinfo[2]{#2}
\providecommand\natexlab[1]{#1}
\providecommand\showeprint[2][]{arXiv:#2}

\bibitem[Agarwal et~al\mbox{.}(2024)]%
        {agarwal2024gkd}
\bibfield{author}{\bibinfo{person}{Rishabh Agarwal}, \bibinfo{person}{Nino
  Vieillard}, \bibinfo{person}{Yongchao Zhou}, \bibinfo{person}{Piotr
  Stanczyk}, \bibinfo{person}{Sabela~Ramos Garea}, \bibinfo{person}{Matthieu
  Geist}, {and} \bibinfo{person}{Olivier Bachem}.}
  \bibinfo{year}{2024}\natexlab{}.
\newblock \showarticletitle{On-Policy Distillation of Language Models: Learning
  from Self-Generated Mistakes}. In \bibinfo{booktitle}{\emph{The Twelfth
  International Conference on Learning Representations}}.
\newblock
\urldef\tempurl%
\url{https://openreview.net/forum?id=3zKtaqxLhW}
\showURL{%
\tempurl}


\bibitem[Aichberger et~al\mbox{.}(2026)]%
        {aichberger2026rethinking}
\bibfield{author}{\bibinfo{person}{Lukas Aichberger}, \bibinfo{person}{Kajetan
  Schweighofer}, {and} \bibinfo{person}{Sepp Hochreiter}.}
  \bibinfo{year}{2026}\natexlab{}.
\newblock \showarticletitle{Rethinking Uncertainty Estimation in {LLM}s: A
  Principled Single-Sequence Measure}. In \bibinfo{booktitle}{\emph{The
  Fourteenth International Conference on Learning Representations}}.
\newblock
\urldef\tempurl%
\url{https://arxiv.org/abs/2412.15176}
\showURL{%
\tempurl}


\bibitem[Chen et~al\mbox{.}(2026)]%
        {chen2026know}
\bibfield{author}{\bibinfo{person}{Hao Chen}, \bibinfo{person}{Ye He},
  \bibinfo{person}{Yuchun Fan}, \bibinfo{person}{Yukun Yan},
  \bibinfo{person}{Zhenghao Liu}, \bibinfo{person}{Qingfu Zhu},
  \bibinfo{person}{Maosong Sun}, {and} \bibinfo{person}{Wanxiang Che}.}
  \bibinfo{year}{2026}\natexlab{}.
\newblock \showarticletitle{Know More, Know Clearer: A Meta-Cognitive Framework
  for Knowledge Augmentation in Large Language Models}.
\newblock \bibinfo{journal}{\emph{arXiv preprint arXiv:2602.12996}}
  (\bibinfo{year}{2026}).
\newblock
\urldef\tempurl%
\url{https://arxiv.org/abs/2602.12996}
\showURL{%
\tempurl}


\bibitem[Choudhury and Sodhi(2024)]%
        {choudhury2024leap}
\bibfield{author}{\bibinfo{person}{Sanjiban Choudhury} {and}
  \bibinfo{person}{Paloma Sodhi}.} \bibinfo{year}{2024}\natexlab{}.
\newblock \showarticletitle{Better than Your Teacher: {LLM} Agents that Learn
  from Privileged {AI} Feedback}.
\newblock \bibinfo{journal}{\emph{arXiv preprint arXiv:2410.05434}}
  (\bibinfo{year}{2024}).
\newblock
\urldef\tempurl%
\url{https://arxiv.org/abs/2410.05434}
\showURL{%
\tempurl}


\bibitem[Feng et~al\mbox{.}(2025)]%
        {feng2025gigpo}
\bibfield{author}{\bibinfo{person}{Lang Feng}, \bibinfo{person}{Zhenghai Xue},
  \bibinfo{person}{Tingcong Liu}, {and} \bibinfo{person}{Bo An}.}
  \bibinfo{year}{2025}\natexlab{}.
\newblock \showarticletitle{Group-in-Group Policy Optimization for {LLM} Agent
  Training}. In \bibinfo{booktitle}{\emph{Advances in Neural Information
  Processing Systems}}, Vol.~\bibinfo{volume}{38}.
\newblock
\urldef\tempurl%
\url{https://proceedings.neurips.cc/paper_files/paper/2025/hash/420c9f777c0b4f78d515e53cf74d58b2-Abstract-Conference.html}
\showURL{%
\tempurl}


\bibitem[Feng et~al\mbox{.}(2026)]%
        {feng2026kbsd}
\bibfield{author}{\bibinfo{person}{Tao Feng}, \bibinfo{person}{Xinke Jiang},
  {and} \bibinfo{person}{Chao Wu}.} \bibinfo{year}{2026}\natexlab{}.
\newblock \showarticletitle{{KbSD}: Knowledge Boundary aware Self-Distillation
  for Behavioral Calibration in Agentic Search}.
\newblock \bibinfo{journal}{\emph{arXiv preprint arXiv:2606.29863}}
  (\bibinfo{year}{2026}).
\newblock
\urldef\tempurl%
\url{https://arxiv.org/abs/2606.29863}
\showURL{%
\tempurl}


\bibitem[Ho et~al\mbox{.}(2020)]%
        {ho2020constructing}
\bibfield{author}{\bibinfo{person}{Xanh Ho}, \bibinfo{person}{Anh-Khoa~Duong
  Nguyen}, \bibinfo{person}{Saku Sugawara}, {and} \bibinfo{person}{Akiko
  Aizawa}.} \bibinfo{year}{2020}\natexlab{}.
\newblock \showarticletitle{Constructing A Multi-hop {QA} Dataset for
  Comprehensive Evaluation of Reasoning Steps}. In
  \bibinfo{booktitle}{\emph{Proceedings of the 28th International Conference on
  Computational Linguistics}}. \bibinfo{publisher}{International Committee on
  Computational Linguistics}, \bibinfo{address}{Barcelona, Spain (Online)},
  \bibinfo{pages}{6609--6625}.
\newblock
\href{https://doi.org/10.18653/v1/2020.coling-main.580}{doi:\nolinkurl{10.18653/v1/2020.coling-main.580}}


\bibitem[Jin et~al\mbox{.}(2025)]%
        {jin2025searchr1}
\bibfield{author}{\bibinfo{person}{Bowen Jin}, \bibinfo{person}{Hansi Zeng},
  \bibinfo{person}{Zhenrui Yue}, \bibinfo{person}{Jinsung Yoon},
  \bibinfo{person}{Sercan~{\"O}. Ar{\i}k}, \bibinfo{person}{Dong Wang},
  \bibinfo{person}{Hamed Zamani}, {and} \bibinfo{person}{Jiawei Han}.}
  \bibinfo{year}{2025}\natexlab{}.
\newblock \showarticletitle{{Search-R1}: Training {LLMs} to Reason and Leverage
  Search Engines with Reinforcement Learning}. In
  \bibinfo{booktitle}{\emph{Conference on Language Modeling}}.
\newblock
\urldef\tempurl%
\url{https://openreview.net/forum?id=Rwhi91ideu}
\showURL{%
\tempurl}


\bibitem[Joshi et~al\mbox{.}(2017)]%
        {joshi2017triviaqa}
\bibfield{author}{\bibinfo{person}{Mandar Joshi}, \bibinfo{person}{Eunsol
  Choi}, \bibinfo{person}{Daniel Weld}, {and} \bibinfo{person}{Luke
  Zettlemoyer}.} \bibinfo{year}{2017}\natexlab{}.
\newblock \showarticletitle{{TriviaQA}: A Large Scale Distantly Supervised
  Challenge Dataset for Reading Comprehension}. In
  \bibinfo{booktitle}{\emph{Proceedings of the 55th Annual Meeting of the
  Association for Computational Linguistics (Volume 1: Long Papers)}}.
  \bibinfo{publisher}{Association for Computational Linguistics},
  \bibinfo{address}{Vancouver, Canada}, \bibinfo{pages}{1601--1611}.
\newblock
\href{https://doi.org/10.18653/v1/P17-1147}{doi:\nolinkurl{10.18653/v1/P17-1147}}


\bibitem[Kwiatkowski et~al\mbox{.}(2019)]%
        {kwiatkowski2019natural}
\bibfield{author}{\bibinfo{person}{Tom Kwiatkowski},
  \bibinfo{person}{Jennimaria Palomaki}, \bibinfo{person}{Olivia Redfield},
  \bibinfo{person}{Michael Collins}, \bibinfo{person}{Ankur Parikh},
  \bibinfo{person}{Chris Alberti}, \bibinfo{person}{Danielle Epstein},
  \bibinfo{person}{Illia Polosukhin}, \bibinfo{person}{Jacob Devlin},
  \bibinfo{person}{Kenton Lee}, \bibinfo{person}{Kristina Toutanova},
  \bibinfo{person}{Llion Jones}, \bibinfo{person}{Matthew Kelcey},
  \bibinfo{person}{Ming-Wei Chang}, \bibinfo{person}{Andrew~M. Dai},
  \bibinfo{person}{Jakob Uszkoreit}, \bibinfo{person}{Quoc Le}, {and}
  \bibinfo{person}{Slav Petrov}.} \bibinfo{year}{2019}\natexlab{}.
\newblock \showarticletitle{{Natural Questions}: A Benchmark for Question
  Answering Research}.
\newblock \bibinfo{journal}{\emph{Transactions of the Association for
  Computational Linguistics}}  \bibinfo{volume}{7} (\bibinfo{year}{2019}),
  \bibinfo{pages}{453--466}.
\newblock
\href{https://doi.org/10.1162/tacl_a_00276}{doi:\nolinkurl{10.1162/tacl_a_00276}}


\bibitem[Liu et~al\mbox{.}(2026)]%
        {liu2026pica}
\bibfield{author}{\bibinfo{person}{Dongyi Liu}, \bibinfo{person}{Yifan Niu},
  \bibinfo{person}{Qinwen Wang}, \bibinfo{person}{Han Xiao}, {and}
  \bibinfo{person}{Jia Li}.} \bibinfo{year}{2026}\natexlab{}.
\newblock \showarticletitle{{PiCA}: Pivot-Based Credit Assignment for Search
  Agentic Reinforcement Learning}.
\newblock \bibinfo{journal}{\emph{arXiv preprint arXiv:2605.09287}}
  (\bibinfo{year}{2026}).
\newblock
\urldef\tempurl%
\url{https://arxiv.org/abs/2605.09287}
\showURL{%
\tempurl}


\bibitem[Lu et~al\mbox{.}(2026)]%
        {lu2026sdar}
\bibfield{author}{\bibinfo{person}{Zhengxi Lu}, \bibinfo{person}{Zhiyuan Yao},
  \bibinfo{person}{Zhuowen Han}, \bibinfo{person}{Zi-Han Wang},
  \bibinfo{person}{Jinyang Wu}, \bibinfo{person}{Qi Gu},
  \bibinfo{person}{Xunliang Cai}, \bibinfo{person}{Weiming Lu},
  \bibinfo{person}{Jun Xiao}, \bibinfo{person}{Yueting Zhuang}, {and}
  \bibinfo{person}{Yongliang Shen}.} \bibinfo{year}{2026}\natexlab{}.
\newblock \showarticletitle{Self-Distilled Agentic Reinforcement Learning}.
\newblock \bibinfo{journal}{\emph{arXiv preprint arXiv:2605.15155}}
  (\bibinfo{year}{2026}).
\newblock
\urldef\tempurl%
\url{https://arxiv.org/abs/2605.15155}
\showURL{%
\tempurl}


\bibitem[Ma et~al\mbox{.}(2026)]%
        {ma2026sdsearch}
\bibfield{author}{\bibinfo{person}{Yufei Ma}, \bibinfo{person}{Zihan Liang},
  \bibinfo{person}{Ben Chen}, \bibinfo{person}{Zhipeng Qian},
  \bibinfo{person}{Huangyu Dai}, \bibinfo{person}{Lingtao Mao},
  \bibinfo{person}{Xuxin Zhang}, \bibinfo{person}{Chenyi Lei}, {and}
  \bibinfo{person}{Wenwu Ou}.} \bibinfo{year}{2026}\natexlab{}.
\newblock \showarticletitle{{SD-Search}: On-Policy Hindsight Self-Distillation
  for Search-Augmented Reasoning}.
\newblock \bibinfo{journal}{\emph{arXiv preprint arXiv:2605.18299}}
  (\bibinfo{year}{2026}).
\newblock
\urldef\tempurl%
\url{https://arxiv.org/abs/2605.18299}
\showURL{%
\tempurl}


\bibitem[Mallen et~al\mbox{.}(2023)]%
        {mallen2023trust}
\bibfield{author}{\bibinfo{person}{Alex Mallen}, \bibinfo{person}{Akari Asai},
  \bibinfo{person}{Victor Zhong}, \bibinfo{person}{Rajarshi Das},
  \bibinfo{person}{Daniel Khashabi}, {and} \bibinfo{person}{Hannaneh
  Hajishirzi}.} \bibinfo{year}{2023}\natexlab{}.
\newblock \showarticletitle{When Not to Trust Language Models: Investigating
  Effectiveness of Parametric and Non-Parametric Memories}. In
  \bibinfo{booktitle}{\emph{Proceedings of the 61st Annual Meeting of the
  Association for Computational Linguistics (Volume 1: Long Papers)}}.
  \bibinfo{publisher}{Association for Computational Linguistics},
  \bibinfo{address}{Toronto, Canada}, \bibinfo{pages}{9802--9822}.
\newblock
\href{https://doi.org/10.18653/v1/2023.acl-long.546}{doi:\nolinkurl{10.18653/v1/2023.acl-long.546}}


\bibitem[{Meta}(2024)]%
        {meta2024llama32}
\bibfield{author}{\bibinfo{person}{{Meta}}.} \bibinfo{year}{2024}\natexlab{}.
\newblock \bibinfo{title}{{Llama 3.2 3B Instruct}: Model Card}.
\newblock \bibinfo{howpublished}{Hugging Face}.
\newblock
\urldef\tempurl%
\url{https://huggingface.co/meta-llama/Llama-3.2-3B-Instruct}
\showURL{%
\tempurl}


\bibitem[Penaloza et~al\mbox{.}(2026)]%
        {penaloza2026privileged}
\bibfield{author}{\bibinfo{person}{Emiliano Penaloza}, \bibinfo{person}{Dheeraj
  Vattikonda}, \bibinfo{person}{Nicolas Gontier}, \bibinfo{person}{Alexandre
  Lacoste}, \bibinfo{person}{Laurent Charlin}, {and} \bibinfo{person}{Massimo
  Caccia}.} \bibinfo{year}{2026}\natexlab{}.
\newblock \showarticletitle{Privileged Information Distillation for Language
  Models}.
\newblock \bibinfo{journal}{\emph{arXiv preprint arXiv:2602.04942}}
  (\bibinfo{year}{2026}).
\newblock
\urldef\tempurl%
\url{https://arxiv.org/abs/2602.04942}
\showURL{%
\tempurl}


\bibitem[Press et~al\mbox{.}(2023)]%
        {press2023measuring}
\bibfield{author}{\bibinfo{person}{Ofir Press}, \bibinfo{person}{Muru Zhang},
  \bibinfo{person}{Sewon Min}, \bibinfo{person}{Ludwig Schmidt},
  \bibinfo{person}{Noah Smith}, {and} \bibinfo{person}{Mike Lewis}.}
  \bibinfo{year}{2023}\natexlab{}.
\newblock \showarticletitle{Measuring and Narrowing the Compositionality Gap in
  Language Models}. In \bibinfo{booktitle}{\emph{Findings of the Association
  for Computational Linguistics: EMNLP 2023}}. \bibinfo{publisher}{Association
  for Computational Linguistics}, \bibinfo{address}{Singapore},
  \bibinfo{pages}{5687--5711}.
\newblock
\href{https://doi.org/10.18653/v1/2023.findings-emnlp.378}{doi:\nolinkurl{10.18653/v1/2023.findings-emnlp.378}}


\bibitem[{Qwen}(2026)]%
        {qwen2026flash}
\bibfield{author}{\bibinfo{person}{{Qwen}}.} \bibinfo{year}{2026}\natexlab{}.
\newblock \bibinfo{title}{{Qwen3.7-Flash}}.
\newblock \bibinfo{howpublished}{QwenCloud model documentation}.
\newblock
\urldef\tempurl%
\url{https://www.qwencloud.com/models/qwen3.7-flash}
\showURL{%
\tempurl}


\bibitem[{Qwen Team}(2024)]%
        {qwen2024qwen25}
\bibfield{author}{\bibinfo{person}{{Qwen Team}}.}
  \bibinfo{year}{2024}\natexlab{}.
\newblock \showarticletitle{{Qwen2.5} Technical Report}.
\newblock \bibinfo{journal}{\emph{arXiv preprint arXiv:2412.15115}}
  (\bibinfo{year}{2024}).
\newblock
\urldef\tempurl%
\url{https://arxiv.org/abs/2412.15115}
\showURL{%
\tempurl}


\bibitem[Ross et~al\mbox{.}(2011)]%
        {ross2011dagger}
\bibfield{author}{\bibinfo{person}{Stephane Ross}, \bibinfo{person}{Geoffrey
  Gordon}, {and} \bibinfo{person}{Drew Bagnell}.}
  \bibinfo{year}{2011}\natexlab{}.
\newblock \showarticletitle{A Reduction of Imitation Learning and Structured
  Prediction to No-Regret Online Learning}. In
  \bibinfo{booktitle}{\emph{Proceedings of the Fourteenth International
  Conference on Artificial Intelligence and Statistics}}
  \emph{(\bibinfo{series}{Proceedings of Machine Learning Research},
  Vol.~\bibinfo{volume}{15})}. \bibinfo{pages}{627--635}.
\newblock
\urldef\tempurl%
\url{https://proceedings.mlr.press/v15/ross11a.html}
\showURL{%
\tempurl}


\bibitem[Schulman et~al\mbox{.}(2017)]%
        {schulman2017ppo}
\bibfield{author}{\bibinfo{person}{John Schulman}, \bibinfo{person}{Filip
  Wolski}, \bibinfo{person}{Prafulla Dhariwal}, \bibinfo{person}{Alec Radford},
  {and} \bibinfo{person}{Oleg Klimov}.} \bibinfo{year}{2017}\natexlab{}.
\newblock \showarticletitle{Proximal Policy Optimization Algorithms}.
\newblock \bibinfo{journal}{\emph{arXiv preprint arXiv:1707.06347}}
  (\bibinfo{year}{2017}).
\newblock
\urldef\tempurl%
\url{https://arxiv.org/abs/1707.06347}
\showURL{%
\tempurl}


\bibitem[Shao et~al\mbox{.}(2024)]%
        {shao2024deepseekmath}
\bibfield{author}{\bibinfo{person}{Zhihong Shao}, \bibinfo{person}{Peiyi Wang},
  \bibinfo{person}{Qihao Zhu}, \bibinfo{person}{Runxin Xu},
  \bibinfo{person}{Junxiao Song}, \bibinfo{person}{Xiao Bi},
  \bibinfo{person}{Haowei Zhang}, \bibinfo{person}{Mingchuan Zhang},
  \bibinfo{person}{Y.~K. Li}, \bibinfo{person}{Y. Wu}, {and}
  \bibinfo{person}{Daya Guo}.} \bibinfo{year}{2024}\natexlab{}.
\newblock \showarticletitle{{DeepSeekMath}: Pushing the Limits of Mathematical
  Reasoning in Open Language Models}.
\newblock \bibinfo{journal}{\emph{arXiv preprint arXiv:2402.03300}}
  (\bibinfo{year}{2024}).
\newblock
\urldef\tempurl%
\url{https://arxiv.org/abs/2402.03300}
\showURL{%
\tempurl}


\bibitem[Shi et~al\mbox{.}(2025)]%
        {shi2025autorefine}
\bibfield{author}{\bibinfo{person}{Yaorui Shi}, \bibinfo{person}{Sihang Li},
  \bibinfo{person}{Chang Wu}, \bibinfo{person}{Zhiyuan Liu},
  \bibinfo{person}{Junfeng Fang}, \bibinfo{person}{Hengxing Cai},
  \bibinfo{person}{An Zhang}, {and} \bibinfo{person}{Xiang Wang}.}
  \bibinfo{year}{2025}\natexlab{}.
\newblock \showarticletitle{Search and Refine During Think: Facilitating
  Knowledge Refinement for Improved Retrieval-Augmented Reasoning}. In
  \bibinfo{booktitle}{\emph{Advances in Neural Information Processing
  Systems}}, Vol.~\bibinfo{volume}{38}.
\newblock
\urldef\tempurl%
\url{https://proceedings.neurips.cc/paper_files/paper/2025/hash/e4a7f5019f06d997e7e8936fcc948c85-Abstract-Conference.html}
\showURL{%
\tempurl}


\bibitem[Su et~al\mbox{.}(2026)]%
        {su2026psp}
\bibfield{author}{\bibinfo{person}{Xingyu Su}, \bibinfo{person}{Abhishek
  Kumar}, \bibinfo{person}{Qing Ping}, \bibinfo{person}{Youzhi Luo},
  \bibinfo{person}{Jonathan Buck}, \bibinfo{person}{Zach Zhang},
  \bibinfo{person}{Subramanian Chidambaram}, {and} \bibinfo{person}{Vinayak
  Arannil}.} \bibinfo{year}{2026}\natexlab{}.
\newblock \showarticletitle{From Self-Distillation to Self-Practice: Privileged
  Information for Multi-Turn Agents}.
\newblock \bibinfo{journal}{\emph{arXiv preprint arXiv:2609.29051}}
  (\bibinfo{year}{2026}).
\newblock
\urldef\tempurl%
\url{https://arxiv.org/abs/2609.29051}
\showURL{%
\tempurl}


\bibitem[Sun et~al\mbox{.}(2026)]%
        {sun2026gapd}
\bibfield{author}{\bibinfo{person}{Xin Sun}, \bibinfo{person}{Jianan Xie},
  \bibinfo{person}{Zhongqi Chen}, \bibinfo{person}{Qiang Liu},
  \bibinfo{person}{Shu Wu}, \bibinfo{person}{Bowen Song},
  \bibinfo{person}{Weiqiang Wang}, \bibinfo{person}{Zilei Wang}, {and}
  \bibinfo{person}{Liang Wang}.} \bibinfo{year}{2026}\natexlab{}.
\newblock \showarticletitle{{GAPD}: Gold-Action Policy Distillation for Agentic
  Reinforcement Learning in Knowledge Base Question Answering}.
\newblock \bibinfo{journal}{\emph{arXiv preprint arXiv:2605.29584}}
  (\bibinfo{year}{2026}).
\newblock
\urldef\tempurl%
\url{https://arxiv.org/abs/2605.29584}
\showURL{%
\tempurl}


\bibitem[Tao et~al\mbox{.}(2026)]%
        {tao2026race}
\bibfield{author}{\bibinfo{person}{Eric Tao}, \bibinfo{person}{Minhao Sun},
  \bibinfo{person}{Liyan Liu}, \bibinfo{person}{Zhijing Wu},
  \bibinfo{person}{Xiaohan Yu}, \bibinfo{person}{Lang Mei}, \bibinfo{person}{Li
  Huang}, \bibinfo{person}{Chao Feng}, \bibinfo{person}{Jinchao Ma},
  \bibinfo{person}{Sichen Kang}, \bibinfo{person}{Siyu Mo}, {and}
  \bibinfo{person}{Chong Chen}.} \bibinfo{year}{2026}\natexlab{}.
\newblock \showarticletitle{{RACE}: Learning Multimodal Search through
  Referential Abstraction and Compositional Evidence}.
\newblock \bibinfo{journal}{\emph{arXiv preprint arXiv:2610.28382}}
  (\bibinfo{year}{2026}).
\newblock
\urldef\tempurl%
\url{https://arxiv.org/abs/2610.28382}
\showURL{%
\tempurl}


\bibitem[Trivedi et~al\mbox{.}(2022)]%
        {trivedi2022musique}
\bibfield{author}{\bibinfo{person}{Harsh Trivedi}, \bibinfo{person}{Niranjan
  Balasubramanian}, \bibinfo{person}{Tushar Khot}, {and}
  \bibinfo{person}{Ashish Sabharwal}.} \bibinfo{year}{2022}\natexlab{}.
\newblock \showarticletitle{{MuSiQue}: Multihop Questions via Single-hop
  Question Composition}.
\newblock \bibinfo{journal}{\emph{Transactions of the Association for
  Computational Linguistics}}  \bibinfo{volume}{10} (\bibinfo{year}{2022}),
  \bibinfo{pages}{539--554}.
\newblock
\href{https://doi.org/10.1162/tacl_a_00475}{doi:\nolinkurl{10.1162/tacl_a_00475}}


\bibitem[Wang et~al\mbox{.}(2026a)]%
        {wang2026igpo}
\bibfield{author}{\bibinfo{person}{Guoqing Wang}, \bibinfo{person}{Sunhao Dai},
  \bibinfo{person}{Guangze Ye}, \bibinfo{person}{Zeyu Gan},
  \bibinfo{person}{Wei Yao}, \bibinfo{person}{Yong Deng},
  \bibinfo{person}{Xiaofeng Wu}, {and} \bibinfo{person}{Zhenzhe Ying}.}
  \bibinfo{year}{2026}\natexlab{a}.
\newblock \showarticletitle{Information Gain-based Policy Optimization: A
  Simple and Effective Approach for Multi-Turn Search Agents}. In
  \bibinfo{booktitle}{\emph{International Conference on Learning
  Representations}}.
\newblock
\urldef\tempurl%
\url{https://openreview.net/forum?id=qkWP6phrvZ}
\showURL{%
\tempurl}


\bibitem[Wang et~al\mbox{.}(2026b)]%
        {wang2026skillsd}
\bibfield{author}{\bibinfo{person}{Hao Wang}, \bibinfo{person}{Guozhi Wang},
  \bibinfo{person}{Han Xiao}, \bibinfo{person}{Yufeng Zhou},
  \bibinfo{person}{Yue Pan}, \bibinfo{person}{Jichao Wang}, \bibinfo{person}{Ke
  Xu}, \bibinfo{person}{Yafei Wen}, \bibinfo{person}{Xiaohu Ruan},
  \bibinfo{person}{Xiaoxin Chen}, {and} \bibinfo{person}{Honggang Qi}.}
  \bibinfo{year}{2026}\natexlab{b}.
\newblock \showarticletitle{{Skill-SD}: Skill-Conditioned Self-Distillation for
  Multi-turn {LLM} Agents}.
\newblock \bibinfo{journal}{\emph{arXiv preprint arXiv:2604.10674}}
  (\bibinfo{year}{2026}).
\newblock
\urldef\tempurl%
\url{https://arxiv.org/abs/2604.10674}
\showURL{%
\tempurl}


\bibitem[Wu et~al\mbox{.}(2026)]%
        {wu2026hiprag}
\bibfield{author}{\bibinfo{person}{Peilin Wu}, \bibinfo{person}{Mian Zhang},
  \bibinfo{person}{Kun Wan}, \bibinfo{person}{Wentian Zhao},
  \bibinfo{person}{Kaiyu He}, \bibinfo{person}{Xinya Du}, {and}
  \bibinfo{person}{Zhiyu Chen}.} \bibinfo{year}{2026}\natexlab{}.
\newblock \showarticletitle{{HiPRAG}: Hierarchical Process Rewards for
  Efficient Agentic Retrieval Augmented Generation}. In
  \bibinfo{booktitle}{\emph{The Fourteenth International Conference on Learning
  Representations}}.
\newblock
\urldef\tempurl%
\url{https://openreview.net/forum?id=Gt4v9WBPzm}
\showURL{%
\tempurl}


\bibitem[Xie et~al\mbox{.}(2026a)]%
        {xie2026evisd}
\bibfield{author}{\bibinfo{person}{Jianan Xie}, \bibinfo{person}{Xin Sun},
  \bibinfo{person}{Zhongqi Chen}, \bibinfo{person}{Xing Zheng},
  \bibinfo{person}{Shu Wu}, \bibinfo{person}{Bowen Song}, {and}
  \bibinfo{person}{Liang Wang}.} \bibinfo{year}{2026}\natexlab{a}.
\newblock \showarticletitle{{EviSD}: Evidence-Conditioned Self-Distillation for
  Search-Augmented Agents}.
\newblock \bibinfo{journal}{\emph{arXiv preprint arXiv:2608.01359}}
  (\bibinfo{year}{2026}).
\newblock
\urldef\tempurl%
\url{https://arxiv.org/abs/2608.01359}
\showURL{%
\tempurl}


\bibitem[Xie et~al\mbox{.}(2026b)]%
        {xie2026tips}
\bibfield{author}{\bibinfo{person}{Yutao Xie}, \bibinfo{person}{Nathaniel
  Thomas}, \bibinfo{person}{Nicklas Hansen}, \bibinfo{person}{Yang Fu},
  \bibinfo{person}{Li~Erran Li}, {and} \bibinfo{person}{Xiaolong Wang}.}
  \bibinfo{year}{2026}\natexlab{b}.
\newblock \showarticletitle{{TIPS}: Turn-Level Information-Potential Reward
  Shaping for Search-Augmented {LLMs}}. In
  \bibinfo{booktitle}{\emph{International Conference on Learning
  Representations}}.
\newblock
\urldef\tempurl%
\url{https://openreview.net/forum?id=eBMOr6a84z}
\showURL{%
\tempurl}


\bibitem[Yang et~al\mbox{.}(2026)]%
        {yang2026rlsd}
\bibfield{author}{\bibinfo{person}{Chenxu Yang}, \bibinfo{person}{Chuanyu Qin},
  \bibinfo{person}{Qingyi Si}, \bibinfo{person}{Minghui Chen},
  \bibinfo{person}{Naibin Gu}, \bibinfo{person}{Dingyu Yao},
  \bibinfo{person}{Zheng Lin}, \bibinfo{person}{Weiping Wang},
  \bibinfo{person}{Jiaqi Wang}, {and} \bibinfo{person}{Nan Duan}.}
  \bibinfo{year}{2026}\natexlab{}.
\newblock \showarticletitle{Self-Distilled {RLVR}}.
\newblock \bibinfo{journal}{\emph{arXiv preprint arXiv:2604.03128}}
  (\bibinfo{year}{2026}).
\newblock
\urldef\tempurl%
\url{https://arxiv.org/abs/2604.03128}
\showURL{%
\tempurl}


\bibitem[Yang et~al\mbox{.}(2018)]%
        {yang2018hotpotqa}
\bibfield{author}{\bibinfo{person}{Zhilin Yang}, \bibinfo{person}{Peng Qi},
  \bibinfo{person}{Saizheng Zhang}, \bibinfo{person}{Yoshua Bengio},
  \bibinfo{person}{William Cohen}, \bibinfo{person}{Ruslan Salakhutdinov},
  {and} \bibinfo{person}{Christopher~D. Manning}.}
  \bibinfo{year}{2018}\natexlab{}.
\newblock \showarticletitle{{HotpotQA}: A Dataset for Diverse, Explainable
  Multi-hop Question Answering}. In \bibinfo{booktitle}{\emph{Proceedings of
  the 2018 Conference on Empirical Methods in Natural Language Processing}}.
  \bibinfo{publisher}{Association for Computational Linguistics},
  \bibinfo{address}{Brussels, Belgium}, \bibinfo{pages}{2369--2380}.
\newblock
\href{https://doi.org/10.18653/v1/D18-1259}{doi:\nolinkurl{10.18653/v1/D18-1259}}


\bibitem[Yao et~al\mbox{.}(2023)]%
        {yao2023react}
\bibfield{author}{\bibinfo{person}{Shunyu Yao}, \bibinfo{person}{Jeffrey Zhao},
  \bibinfo{person}{Dian Yu}, \bibinfo{person}{Nan Du}, \bibinfo{person}{Izhak
  Shafran}, \bibinfo{person}{Karthik Narasimhan}, {and} \bibinfo{person}{Yuan
  Cao}.} \bibinfo{year}{2023}\natexlab{}.
\newblock \showarticletitle{{ReAct}: Synergizing Reasoning and Acting in
  Language Models}. In \bibinfo{booktitle}{\emph{The Eleventh International
  Conference on Learning Representations}}.
\newblock
\urldef\tempurl%
\url{https://openreview.net/forum?id=WE_vluYUL-X}
\showURL{%
\tempurl}


\bibitem[Yu et~al\mbox{.}(2026)]%
        {yu2026dopd}
\bibfield{author}{\bibinfo{person}{Xinlei Yu}, \bibinfo{person}{Gen Li},
  \bibinfo{person}{Qingyi Si}, \bibinfo{person}{Guibin Zhang},
  \bibinfo{person}{Yuqi Xu}, \bibinfo{person}{Congcong Wang},
  \bibinfo{person}{Shuai Dong}, \bibinfo{person}{Kaiwen Tuo},
  \bibinfo{person}{Xiangyu Zeng}, \bibinfo{person}{Kaituo Feng},
  \bibinfo{person}{Qunzhong Wang}, \bibinfo{person}{Yang Shi},
  \bibinfo{person}{Xiaobin Hu}, \bibinfo{person}{Xiangyu Yue},
  \bibinfo{person}{Jiaqi Wang}, {and} \bibinfo{person}{Shuicheng Yan}.}
  \bibinfo{year}{2026}\natexlab{}.
\newblock \showarticletitle{{DOPD}: Dual On-policy Distillation}.
\newblock \bibinfo{journal}{\emph{arXiv preprint arXiv:2606.30626}}
  (\bibinfo{year}{2026}).
\newblock
\urldef\tempurl%
\url{https://arxiv.org/abs/2606.30626}
\showURL{%
\tempurl}


\bibitem[Zhang(2026)]%
        {zhang2026aropd}
\bibfield{author}{\bibinfo{person}{Wenhao Zhang}.}
  \bibinfo{year}{2026}\natexlab{}.
\newblock \showarticletitle{Beyond Absolute Imitation: Anchored Residual
  Guidance for Privileged On-Policy Distillation}.
\newblock \bibinfo{journal}{\emph{arXiv preprint arXiv:2606.10385}}
  (\bibinfo{year}{2026}).
\newblock
\urldef\tempurl%
\url{https://arxiv.org/abs/2606.10385}
\showURL{%
\tempurl}


\end{thebibliography}
\end{document}